\documentclass{emulateapj}
\usepackage{natbib}
\bibpunct{(}{)}{;}{a}{}{,}
\usepackage{graphicx}
\usepackage{amsmath,amssymb}
\usepackage[dvipdfm,breaklinks=true,bookmarks=true,hypertexnames=false,bookmarksnumbered=true,bookmarksopen=false,bookmarks=true,bookmarksnumbered=true]{hyperref}
\newcommand{\msun}{{\rm M}_{\odot}}

\newcommand{\kms}{\, {\rm km\, s}^{-1}}
\newcommand{\h}{\,h_{70}}
\newcommand{\hmm}{\h^{-1}}
\newcommand{\hmmsun}{\hmm\msun}
\newcommand{\kpc}{\, {\rm kpc}}

\newcommand{\hmkpc}{\hmm\kpc}

\newcommand{\hmmolsun}{\h\, \left(M/L_V\right)_{\odot}}
\newcommand{\Mpc}{\, {\rm Mpc}}

\newcommand{\der}{{\rm d}}

\newcommand{\mypm}[2]{^{+#1}_{-#2}}

\newcommand{\dos}{D_{\rm os}}
\newcommand{\dol}{D_{\rm ol}}
\newcommand{\dls}{D_{\rm ls}}
\newcommand{\dlsdos}{\dls/\dos}
\newcommand{\eg}{{\it e.g.}}
\newcommand{\ie}{{\it i.e.}}

\newcommand{\zl}{z_{\rm l}}
\newcommand{\zs}{z_{\rm s}}
\newcommand{\zsi}{z_{\rm s1}}
\newcommand{\zsii}{z_{\rm s2}}

\newcommand{\rein}{R_{\rm Ein}}
\newcommand{\reff}{R_{\rm eff}}

\newcommand{\lname}{SDSSJ0946+1006}

\newcommand{\blobs}{components}
\newcommand{\noiserms}{0.03}
\begin{document}
\title{The Sloan Lens ACS Survey. VI: Discovery and analysis of a double Einstein ring\altaffilmark{$\dagger$}.}
\author{Rapha\"el Gavazzi\altaffilmark{1}}
\author{Tommaso Treu\altaffilmark{1,2,3}}
\author{L\'eon V. E. Koopmans\altaffilmark{4}}
\author{Adam S. Bolton\altaffilmark{5,6}}
\author{Leonidas A. Moustakas\altaffilmark{7}}
\author{Scott Burles\altaffilmark{8}}
\author{Philip J. Marshall\altaffilmark{1}}

\altaffiltext{$\dagger$}{Based on observations made with the NASA/ESA Hubble
Space Telescope, obtained at the Space Telescope Science Institute,
which is operated by the Association of Universities for Research in
Astronomy, Inc., under NASA contract NAS 5-26555.  These observations
are associated with program \#10886.  Support for program \#10886 was
provided by NASA through a grant from the Space Telescope Science
Institute, which is operated by the Association of Universities for
Research in Astronomy, Inc., under NASA contract NAS 5-26555.}
\altaffiltext{1}{Department of Physics, University of California, Broida Hall, Santa Barbara, CA 93106-9530, USA}
\altaffiltext{2}{Sloan Fellow}
\altaffiltext{3}{Packard Fellow}
\altaffiltext{4}{Kapteyn Astronomical Institute, University of Groningen, PO box 800, 9700 AV Groningen, The Netherlands}
\altaffiltext{5}{Institute for Astronomy, University of Hawaii, 2680 Wodlawn Dr., Honolulu, HI 96822, USA}
\altaffiltext{6}{Harvard-Smithsonian Center for Astrophysics, 60 Garden St. MS-20, Cambridge, MA 02138, USA}
\altaffiltext{7}{Jet Propulsion Laboratory, Caltech, MS 169-327, 4800 Oak Grove Dr., Pasadena, CA 91109, USA}
\altaffiltext{8}{Department of Physics and Kavli Institute for Astrophysics and Space Research, Massachusetts Institute of Technology, 77 Massachusetts Ave., Cambridge, MA 02139, USA}

\email{gavazzi@iap.fr}
\shorttitle{The double Einstein ring \lname}
\shortauthors{Gavazzi et~al.}

%..............................
\begin{abstract}
We report the discovery of two concentric partial Einstein rings around the
gravitational lens \lname, as part of the Sloan Lens ACS Survey. The
main lens is at redshift $\zl=0.222$, while the inner ring (1)
is at redshift $\zsi=0.609$ and Einstein radius ${\rein}_1=1.43\pm0.01 \arcsec$.
The wider image separation (${\rein}_2=2.07\pm 0.02\arcsec$) of the outer ring
(2) implies that it is at higher redshift than Ring 1. Although
no spectroscopic feature was detected in $\sim9$ hours of spectroscopy at
the Keck I Telescope, the detection of Ring 2 in the F814W ACS filter
implies an upper limit on the redshift of $\zsii\lesssim6.9$. The lens
configuration can be well described by a power law total mass density
profile for the main lens $\rho_{\rm tot}\propto r^{-\gamma'}$ with
logarithmic slope $\gamma'=2.00\pm0.03$ (i.e. close to isothermal),
velocity dispersion $\sigma_{\rm SIE}=287\pm5\kms$ (in good agreement with
the stellar velocity dispersion $\sigma_{v,*}=284\pm24\kms$) with little
dependence upon cosmological parameters or the redshift of Ring 2.
Using strong lensing constraints only we show that the enclosed mass to
light ratio increases as a function of radius, inconsistent with mass
following light. Adopting a prior on the stellar mass to light ratio from
previous SLACS work we infer that $73\pm9\%$ of the mass is in form of dark
matter within the cylinder of radius equal to the effective radius of
the lens. We consider whether the double source plane configuration can be used to
constrain cosmological parameters exploiting the ratios of angular
distance ratios entering the set of lens equations. We find that
constraints for \lname\, are uninteresting due to the sub-optimal lens
and source redshifts for this application. We then consider the
perturbing effect of the mass associated with Ring 1 (modeled as a
singular isothermal sphere) building a double lens plane compound lens model.
This introduces minor changes to the mass of the main lens,
allows to estimate the redshift of the Ring 2 $(\zsii=3.1\mypm{2.0}{1.0})$,
and  the mass of the source responsible for Ring 1
$(\sigma_{\rm SIE,s1}=94\mypm{27}{47}\kms)$.
We conclude by examining the prospects of doing cosmography with a sample
of 50 double source plane gravitational lenses, expected from future space based surveys such as
DUNE or JDEM. Taking full account of the uncertainty in the mass density
profile of the main lens, and of the effect of the perturber, and assuming
known redshifts for both sources, we find that such a sample could be used
to measure $\Omega_{\rm m}$ and $w$ with 10\% accuracy, assuming a flat
cosmological model.
\end{abstract}
%\maketitle

%======================================================================
%======================================================================
%======================================================================
\section{Introduction}\label{sec:intro}

Measuring the mass distribution of galaxies is essential for
understanding a variety of astrophysical processes.
Extended mass profiles of galaxies provide evidence for dark matter
either using rotation curves \citep[\eg][]{rubin80,albada85,swaters03}, weak lensing
\citep[\eg][]{brainerd96,hoekstra04,sheldon04,mandelbaum06}, or dynamics of
satellite galaxies \citep[\eg][]{prada03,conroy07} which is one of the main
ingredients of the standard $\Lambda$ cold dark matter ($\Lambda$CDM) cosmological
model. At galactic and subgalactic scales, numerical cosmological simulations
make quantitative predictions regarding, e.g., the inner slope of mass
density profiles and the existence of dark matter substructure.
Precise mass measurements are key to test the
predictions and provide empirical input to further improve the models.

Gravitational lensing has emerged in the last two decades as one of the
most powerful ways to measure the mass distributions of galaxies,
by itself or in combination with other diagnostics. Although strong
gravitational lenses are relatively rare in the sky
\citep[$\lesssim20$ per square degree at space-based depth and
resolution;][]{marshall05,moustakas07}, the number of known galaxy-scale
gravitational lens systems has increased well beyond a hundred as a result
of a number of dedicated efforts exploiting a variety of techniques
\citep[\eg][]{warren96,ratnatunga99,kochanek99,myers03b,bolton04,cabanac07}.
The increased number of systems, together with the improvement of modeling
techniques \citep[\eg][]{kochanek92,warren03,treu04,brewer06,suyu06,wayth06,barnabe07},
has not only enabled considerable progress in the
use of this diagnostic for the study of the mass distribution of early
and most recently late-type galaxies, but also for cosmography,
i.e. the determination of cosmological parameters
\citep[\eg][]{golse02a,soucail04,dalal05}.

Given the already small optical depth for strong lensing, the lensing of multiple
background sources by a single foreground galaxy is an extremely rare event.
At Hubble Space Telescope (HST) resolution (FWHM $\sim0\farcs12$) and
depth (I$_{\rm AB}\sim27$) it is expected that one massive early-type
galaxy (which dominate the lensing cross-section) in about 200 is a strong
lens \citep{marshall05}. Taking into account the strong dependence of the
lensing cross-section on lens galaxy velocity dispersion ($\propto \sigma^4$),
and the population of lens galaxies, we estimate that about one lens galaxy in
$\sim 40-80$ could be a double source plane strong gravitational lens
(see appendix \ref{append:prob}).
For these reasons, at most a handful of double lenses are to be found in the
largest spectroscopic surveys of early-type galaxies such as the luminous
red galaxies of the Sloan Digital Sky Survey. However, future high resolution
imaging surveys such as those planned for JDEM and DUNE
\citep{aldering04,refregier06} will increase the number of known lenses by 2-3
orders of magnitude \citep{marshall05}, and hence should be able to provide large
statistical samples of double source plane gravitational lenses, opening up the
possibility of qualitatively new applications of gravitational lensing for the
study of galaxy formation and cosmography.

We report here the discovery of the first double source plane partial Einstein Ring.
The gravitational lens system \lname~, was discovered as part of the Sloan
Lens ACS (SLACS) Survey
\citep{bolton05a,bolton06,treu06,koopmans06,bolton07a,gavazzi07b}.
The object was first
selected by the presence of multiple emission lines at higher redshift
in the residuals of an absorption line spectrum from the SDSS
database as described by \citet{bolton04} and then confirmed as a
strong lens by high resolution imaging with the Advanced Camera for
Surveys aboard HST. In addition to an Einstein ring due
to the source (hereafter source 1) responsible for the emission lines
detected in the SDSS spectrum, the Hubble image also shows a second
multiply imaged system forming a broken Einstein Ring with a larger
diameter then the inner ring (hereafter source 2). 
This configuration can only arise if the two lensed systems are
at different redshifts and well aligned with the center of
the lensing galaxy. It is a great opportunity that a double source
plane lens has been found among the approximately 90 lenses discovered
by the SLACS collaboration to date \citep[][in prep.]{bolton08}.

The goal of this paper is to study and model this peculiar system in
detail, as an illustration of some astrophysical applications of
double source plane compound lenses, including i) the determination of the mass
density profile of the lens galaxy independent of dynamical constraints;
ii) placing limits on the mass of source 1 based on multiple lens plane 
modeling; iii) estimating the redshift of source 2 and the cosmological
parameters from the angular distance size ratios. The paper is therefore
organized as follows. Section \ref{sec:data} summarizes
the observations, photometric and spectroscopic measurements, and
discusses the morphology of the lens system. Section \ref{sec:model} describes our
gravitational lens modeling methodology. Section \ref{ssec:resodel} gives the main
results in terms of constraints on the mass distribution of the lens
galaxy and of source 1. Section \ref{sec:beyond} discusses the use of double source plane
lenses as a tool for cosmography using the example of \lname~and also
addresses the potential of large samples of such double source plane lenses for
the same purpose. In section \ref{sec:conclu} we summarize our results and briefly conclude.

Unless otherwise stated we assume a concordance cosmology
with $H_0 = 70\h\,\kms\Mpc^{-1}$, $\Omega_{\rm m}=0.3$ and
$\Omega_\Lambda=0.7$. All magnitudes are expressed in the AB system.

%----------------------------------------
\begin{figure*}[htb]
  \centering
  \includegraphics[width=\hsize]{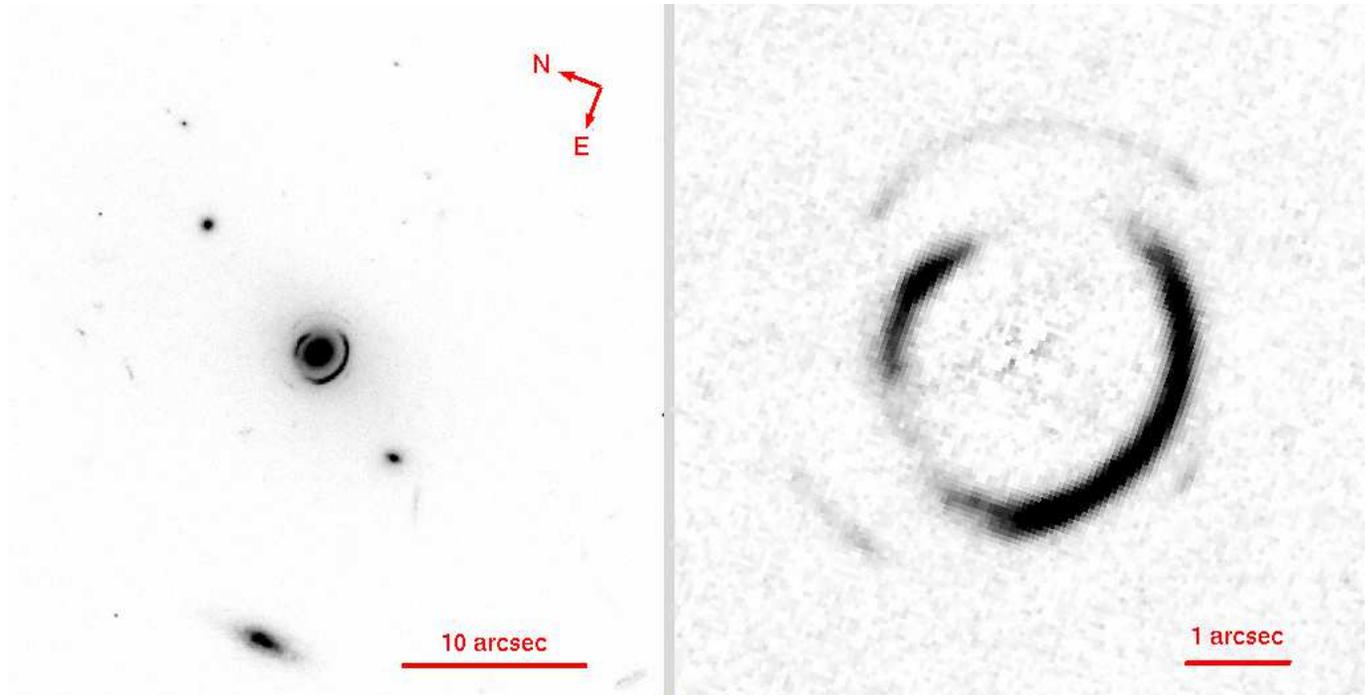}
  \caption{\small HST/F814W overview of the lens system \lname. The right panel is a zoom onto the lens showing two concentric partial ring-like structures after subtracting the lens surface brightness.}
  \label{fig:overview}
\end{figure*}
%======================================================================
%======================================================================
\section{Data}\label{sec:data}

The lens galaxy \lname~was first identified in the spectroscopic SDSS
database based on the redshift of the lensing galaxy $\zl=0.222$ and
that of a background source at $\zsi=0.609$ (hereafter source 1), as
described by \citet{bolton04,bolton06}, and
\citet[][in prep.]{bolton08}. This section describes HST follow-up imaging 
(\S\ref{ssec:obs}), the properties of the lens (\S\ref{ssec:lens-pptes}) 
and lensed (\S\ref{ssec:lensed}) galaxies.
%, and spectroscopic follow-up with the 10m Keck Telescope (\S~\ref{ssec:spectro}).

\subsection{Hubble Space Telescope observations \& data reduction}
\label{ssec:obs}

\lname~was then imaged with the ACS on board the HST
(cycle 15, Prog. 10886, PI Bolton).
The Wide Field Channel with filter F814W was used for a total
exposure time of 2096 s. Four sub-exposures were obtained with a
semi-integer pixel offset ({\tt acs-wfc-dither-box}) to ensure proper
cosmic ray removal and sampling of the point spread function. The
image reduction process is described in \citep{gavazzi07b} and results
in a $0\farcs03$/pixel spatial sampling. This pixel size provides good
sampling of the PSF for weak lensing applications, at the (small) price
of inducing noise correlation over scales of 1-2 pixels. This is
accounted for in our analysis by correcting pixel variances
according to the procedure described by \citet{casertano00}.

Figure \ref{fig:overview} shows the HST image of the lens galaxy field
together with an enlarged view of the lensed features, after
subtraction of a smooth model for the lens surface brightness
distribution.  For reference, one arcsecond in the lens plane subtends
a physical scale of $3.580\hmkpc$.

%------------------------------------------------------------
\subsection{Lens galaxy properties}\label{ssec:lens-pptes}

The two-dimensional lens surface brightness was fitted with {\tt
galfit} \citep{peng02} using two elliptical S\'ersic components. The
addition of a second component is needed to provide a good fit in the
center, and to reproduce the isophotal twist in the outer regions. To
reduce the effect of lensed features in the fit we proceeded
iteratively. We first masked the lensed features manually, then we
performed {\tt galfit} fits creating masks by 4-$\sigma$ clipping.
Two iterations were needed to achieve convergence.

The total magnitude of the lens obtained by summing the flux of the two
S\'ersic models is $F814W = 17.110\pm 0.002$ after correction for Galactic
extinction \citep{schlegel98}. The rest-frame V band absolute magnitude is
$M_V=-22.286 \pm 0.025$ using the K-correction of \citet{treu06}.
The errors are dominated by systematic uncertainties on the K-correction term. The most
concentrated S\'ersic component $c_1$ dominates at the center and
accounts for about 17.5\% of the total lens flux. The effective radius
of $c_1$ is about $0.4\arcsec$ whereas that of $c_2$ is $\sim
3\arcsec$ with about 10\% relative accuracy. Similarly, the S\'ersic
indexes are $n_{c_1}\simeq 1.23$ and $n_{c_2}\simeq1.75$. 

\begin{figure}[htb]
  \centering
  \includegraphics[width=\hsize,height=7cm]{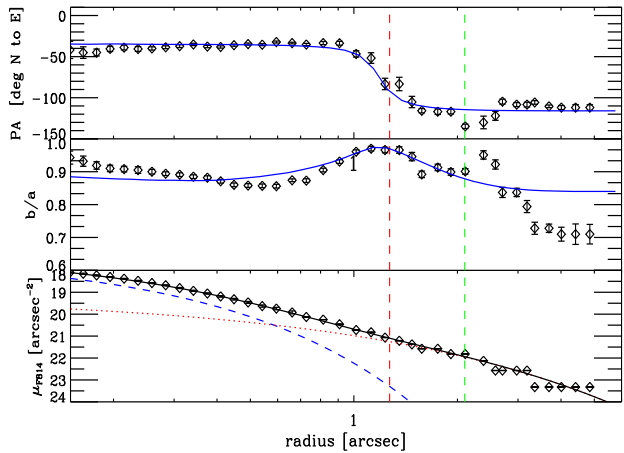}
  \caption{\small Results for isophotal fit with {\tt IRAF/ellipse}. \ \ \ {\it Top panel:} Position angle versus radius. \ \ \ {\it Middle panel:} Axis ratio versus radius.  The vertical lines show the location of the inner and outer Einstein rings which were masked out during the fitting process. We also overlay in the top and middle panels as a blue solid line the {\tt ellipse} output performed on the best fit {\tt galfit} two-dimensional brightness distribution. \ \ \ {\it Bottom panel:} best fit S\'ersic profiles obtained with {\tt galfit}. The formal error bars on the surface brightness profile are smaller than the data points.}
  \label{fig:el-pa-lens}
\end{figure}

To measure the one dimensional light profile of the lens galaxy, we
used the {\tt IRAF} task {\tt ellipse}. Fig. \ref{fig:el-pa-lens}
shows the radial change of ellipticity and position angle of the light
distribution. There is a clear indication of a sharp change in
position angle and ellipticity between $1-2\arcsec$. This isophotal
twist is well captured by the double S\'ersic profile fit, that requires
different PAs for the two components. Therefore we conclude that the
lens galaxy is made of two misaligned components, having similar
surface brightness at radius $\sim 0.6\arcsec$.  

For comparison, a single component S\'ersic fit yields $n\simeq3.73$,
consistent with the typical light profiles of massive early-type galaxies.
The effective radius of the composite surface brightness distribution is
found to be $\reff=2.02 \pm 0.10 \ {\rm arcsec}\simeq 7.29\pm0.37
\hmkpc$, where we assumed a typical relative uncertainty of about
5\% as discussed in \citep{treu06}. It is also consistent with an
independent measurement reported by \citet[][in prep.]{bolton08} who considered 
de Vaucouleurs surface brightness distributions ($n\equiv 4$ by construction).
Note that we use the same convention for all characteristic radii reported
throughout. For elliptical distributions radii are expressed at the intermediate
radius  (\ie~the geometric mean radius $r=\sqrt{a b}$).

In addition, the stellar velocity dispersion $\sigma_{\rm ap}=263 \pm 21 \kms$
was measured with  SDSS spectroscopy within a $3\arcsec$ diameter fiber.
We convert this velocity dispersion $\sigma_{\rm ap}$ into the fiducial
velocity $\sigma_{v,*}$ that enters Fundamental Plane analyses and measured
in an aperture of size $\reff/8$  using the relation
$\sigma_{v,*}/\sigma_{\rm ap}=(\reff/8/R_{\rm ap})^{-0.04}\simeq 1.08$
\citep[see][and references therein]{treu06}

Based on photometric redshifts available online on the SDSS webpage
\citep{oyaizu07}, we note that the lens galaxy is the brightest galaxy
in its neighborhood. Another bright galaxy about 40 arcsec south-west
of \lname\, exhibits perturbed isophotes (an extended plume) suggesting
that it may have flown by recently and might end up merging onto the
lens galaxy. Its photometric redshift is $z_{\rm phot}=0.20\pm0.04$
consistent with \lname~redshift. The extended envelope captured by the double
S\'ersic component fit also supports the recent flyby hypothesis
\citep[\eg~][]{bell06a}.

%----------------------------------------

%------------------------------------------------------------
\subsection{Lensed structures}\label{ssec:lensed}
Two concentric partial ring-like structures are clearly seen at radii
$1.43\pm0.01\arcsec$ and $ 2.07\pm 0.02\arcsec$ from the center
of the lens galaxy (Figure~\ref{fig:overview}). Such a peculiar lensing
configuration -- with widely different image separations of nearly concentric
multiple image systems -- implies that the rings come from two sources
at different redshift, the innermost (Ring 1) corresponding to the
nearest background source 1 and the outermost (Ring 2) being
significantly further away along the optical axis.

Ring 1 has a typical cusp configuration with 3 merging conjugate images
and a counterimage on the opposite side of the lens and closer to the center than the large cusp ``arc''. This constrains the
orientation of the lens potential major axis to pass almost through the middle
of both arcs. Ring 1 is among the brightest ones to have been discovered
in the SLACS survey \citep[See][in prep., for the latest compilation]{bolton08}.
The observed F814W magnitude is
$m_{\rm 1}=19.784 \pm 0.006$ (extinction corrected). The error bar includes 
only statistical uncertainties. An additional systematic error of order
$\lesssim0.1$ mag is likely present due to uncertainties in the lens galaxy
subtraction \citep{marshall07}.

Ring 2 presents a nearly symmetrical Einstein cross configuration
(with a faint bridge between the north and west images), implying that
the source must lie very close to the optical axis. The observed F814W
magnitude is $m_{\rm 2}=23.68 \pm 0.09$, making it about 36 times
fainter than Ring 1. As for Ring 1 the error bar includes only
statistical uncertainties.

No evidence of Ring 2 is present in the SDSS spectrum.
This can be explained by the low peak surface brightness of Ring 2
($\sim 23\ {\rm mag\ arcsec}^{-2}$)  and less importantly by the diameter
of the second ring being slighty larger than the $3\arcsec$ SDSS fiber
\citep[although see][for a successful redshift measurement in a similar case]{bolton06b}.
Deeper longslit spectroscopy was obtained at Keck Observatory with the Low Resolution Imager Spectrograph (LRIS)
instrument on December 22-23 2006, the total integration time being about
9 hours. The goal was two-fold: i) obtain the redshift of Ring 2;
 ii) measure the stellar velocity dispersion profile of the main lens. 
This latter aspect will be presented elsewhere. Despite the large
integration time, we could not measure the source redshift  $\zsii$
due to a lack of emission lines in the range [3500, 8600\AA] that do not
belong to Ring 1. Since Ring 2 is detected in the ACS/F814W filter, we can
set an upper limit on its redshift $\zsii<6.9$ by requiring that the
Lyman break be at shorter wavelengths than the red cutoff of the 
filter.

%======================================================================
%======================================================================
\section{Lens modeling}\label{sec:model}

%------------------------------------------------------------
\subsection{Model definition}\label{ssec:methodel}

This section describes our adopted strategy to model this exceptional
lens system. We begin with a simplifying assumption. Although the
gravitational potential arises from both a stellar and a dark matter
component,  a single power law model for the total density profile turns out
to be a good description of SLACS lenses
\citep{koopmans06}. Therefore, we assume the total convergence for a
source at redshift $\zs$ to be of the form:
\begin{equation}\label{eq:densprof}
  \kappa(\vec{r},\zs) = \frac{b_\infty^{\gamma'-1}}{2} \left( x^2 + y^2/q^2
  \right)^{(1-\gamma')/2} \frac{\dls}{\dos} \;,
\end{equation}
with 4 free parameters: the overall normalization $b$, the logarithmic
slope of the density profile $\gamma'$, the axis ratio $q$ and
position angle $PA_0$ (omitted in Eq. \eqref{eq:densprof} for simplicity)
of iso-$\kappa$ ellipses. The familiar case of the singular isothermal
sphere is that corresponding to a slope $\gamma'=2$ and $q=1$. In this
case $b_\infty$ relates to the velocity dispersion of the isothermal profile by
$b_\infty = 4 \pi (\sigma_{\rm SIE}/c)^2 = ( \sigma_{\rm SIE} /  186.2 \kms)^2 \ {\rm arcsec}$.
Note that $\sigma_{\rm SIE}$ is nothing but a way of redefining
the normalization of the convergence profile and does not necessarily
correspond in a straightforward sense to the velocity dispersion   
of stars in the lens galaxy. In general, for every combination of 
model parameters, the stellar velocity dispersion of a specified
tracer embedded in the potential can be computed by solving the Jeans
equation and will be a function of radius and observational effects
such as aperture and seeing.

No assumptions are made about the orientation of the position angle
$PA_0$ of the lens potential. In addition, we allow for external shear
with modulus $\gamma_{\rm ext}$ and position angle $PA_{\rm ext}$.
For a multiple source plane system, it is necessary to define a lens plane
  from which the external shear comes from since shear has to be scaled by the
  apropriate $\dls/\dos$ term for each source plane. For simplicity we assume
  that the global effect of external pertubations comes from the same lens
  plane $\zl=0.222$.
We expect a strong degeneracy between internal ellipticity and external
shear but include this extra degree of freedom in the model to account
for any putative twist of isopotentials, as suggested by the observed
isophotal twist in the lens galaxy surface brightness. Note also that
the need of being able to handle two distinct source planes led us to the
somewhat unusual definition of $b_\infty$ in Eq. \eqref{eq:densprof}. With
this convention, $ (b_\infty \sqrt{q})^{\gamma'-1} \dlsdos$ is the quantity
closest to the $b_{\rm SIE}$ (or $R_{\rm Einst}$) parameter used in other SLACS
papers \citep{koopmans06,bolton08}. Note also that the center of mass is
assumed to match exactly the lens galaxy center of light. The unknown redshift
of source 2 is also treated as a free parameter, for which we assign a uniform
$1\le\zsii\le6.9$ prior. Altogether, we use 7 free parameters to characterize
the~potential of~\lname: $b_\infty$, $\gamma'$, $q$, $PA_0$, $\gamma_{\rm ext}$,
$PA_{\rm ext}$ and $\zsii$.

In this section and the next, we neglect the extra focusing effect of
Ring 1 acting as a lens on Ring 2, leaving the discussion of this
perturber for Section~\ref{sec:beyond}.

%------------------------------------------------------------
\subsection{Methods}\label{ssec:methods}
We consider three strategies for studying  gravitational lens systems with
spatially resolved multiple images.

The first one consists of identifying conjugate bright spots in
the multiple images and minimizing the distance of conjugate points in
the source plane. This approach is statistically conservative in the
sense that it only takes partial advantage of the large amount of
information present in the deep HST data. However, it has the benefit
of being robust and relatively insensitive to the details of the
source morphology, and other concerns that affect different
alternative techniques in the case.

The second approach is the linear source inversion and parametric
potential fitting method described by \citet{warren03},
\citet{treu04}, \citet{Koopmans05} and \citet{suyu06}. A strong advantage of 
this method is that it takes fully into account the amount of
information contained in each pixel. Although this method is robust,
there are many degrees of freedom to model the intrinsic source
surface brightness distribution and thus some form of regularisation
is needed to avoid fitting the noise as described in the references
above.

The third method \citep[\eg][]{marshall07,bolton07a,bolton08} describes the
source as one or several \blobs~parameterized with elliptical surface
brightness profiles (usually S\'ersic). In general, this
method provides good fits to the data as long as not too many such
\blobs~are needed to represent the source, and directly provides 
physically meaningful parameters for the source. For high signal-to-noise
ratio images of complex lensed features the dimensionality of the
problem may increase very fast.

In the case of a multiple source plane system, two difficulties arise when using the
second and third techniques. 1) Our current pixellized method does not handle multiple
source planes \citep[see \eg][for recent progress along this line]{dye07}.
2) The statistical weight given to each of the partial rings depends
essentially on their relative brightness. Since Ring 1 is 36 times brighter
than Ring 2, it completely dominates the fit. This has the unwanted side effect
that a physically uninteresting morphological mismatch of the inner ring, due
for example to poor modeling of the source or of the point spread function,
overwhelms any mismatch in the physically important {\it image separation}
of the outer ring.

The goal of the present analysis is to confirm that \lname~is the first
example of a galaxy-scale double source plane  system and illustrate what kind of
information can be inferred from such a  configuration.
After experimenting with all three techniques -- and in light of the
difficulties described above --  we decided to focus on the more
straightforward conjugate points modeling technique, using the other
techniques to aid in our modeling.

In practice, the modeling technique adopted here is similar to
the one used by \citet{gavazzi03}. The merging cusp nature of ring
1 makes the identification of quadruply imaged spots hazardous along
the elongated arc but identifications are much easier between the
opposite counter-image and the elongated arc. The identification
of the brightness peak S2 in Ring 2 is obvious. To guide the
identification process, we also used fits based on the pixellized
source inversion. We ended up having 4 spots identified in Ring 1, two
of them having 3 clear conjugations (S1a, S1c) whereas the other two have only
have 2 (S1b, S1d). One single bright spot in Ring 2 is imaged 4 times. The
typical rms error made on the location of spots estimated to be
$\noiserms\arcsec$.
Table \ref{tab:points} summarizes the coordinates of matched points in the same frame as Fig. \ref{fig:overview}. For each knot S1a, S1b, S1c, S1d and S2, multiple images with positive parity have an odd labelling number. To guide the fitting procedure we also demand the image parity to be preserved by the model. Therefore, taking into account the unknown
position of these spots in the source plane, we end up having  18
%$2\times 2\times(3-1)+ 2 \times 2 \times(2-1) + 2\times (4-1) = 18 $
constraints \citep[see][]{gavazzi03} whereas the considered model has
7 free parameters. Hence the optimization problem has 11 degrees of
freedom.

%.................................
\begin{table}[htbp]
\begin{center}
  \caption{Summary of pixel coordinates used for lens modeling.}\label{tab:points}
  \begin{tabular}{lcccc}\hline\hline
          &  Img. 1         &  Img. 2        &   Img. 3      &  Img. 4    \\\hline
    S1a   &  0.34, -1.50  & -0.94, 0.68  &  1.52, 0.19 &  --        \\
    S1b   &  --           & -1.16, 0.22  &  1.44, 0.88 &  --        \\
    S1c   & -0.43, -1.42  & -1.10, 0.67  &  1.23, 0.88 &  --        \\
    S1d   & -0.14, -1.68  & -0.57, 0.96  &  --         &  --        \\
    S2    & -1.51, -1.78  &  1.56,-1.19  &  1.55, 1.65 & -1.34, 1.32 \\
    \hline\end{tabular}
\end{center}
{\small Positions (x,y) of each multiple knot are expressed in arcsec (typical rms error $0.03\arcsec$) relative to the lens galaxy surface brightness peak (got from {\tt galfit} modeling, see \S\ref{ssec:lens-pptes}). The frame position angle is $161.348^\circ$ relative to North direction. 
%09:46:56.655 +10:06:52.84 
}
\end{table}

%----------------------------------------
\begin{figure*}[htb]
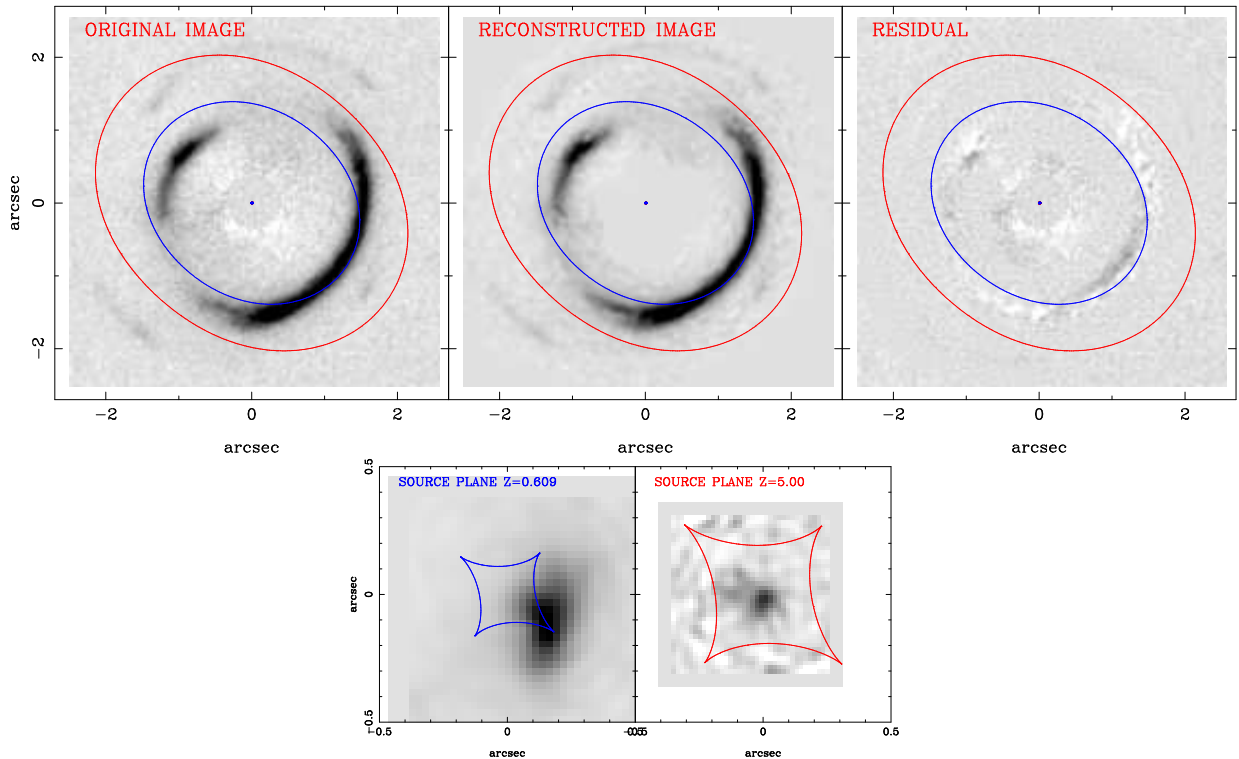

  \centering
  \includegraphics[height=6cm]{best_fit_2src1.eps}
  \includegraphics[height=4cm]{best_fit_2src2.eps}
  \caption{\small Best-fit single lens plane model for the lens \lname. The model parameters were found using the identification of conjugate bright knots but the quality of the model is illustrated with a pixelised source inversion technique. {\it Top left:} observation with the lens light profile subtracted off. {\it Top middle:} model prediction in the image plane and associated residuals ({\it Top right}). The model also predicts the light distribution in the source planes $\zsi$ and $\zsii$ ({\it Bottom left} and {\it right} respectively). Note a different color stretching for source plane 2 (factor 6) in this latter case. Critical and caustic lines corresponding to the two source planes are overlaid (smaller blue for $\zsi=0.609$ and wider red for $\zsii=5$).}
  \label{fig:model1}
\end{figure*}

%======================================================================
%======================================================================
\section{Modeling Results}\label{ssec:resodel}

The optimization process and the exploration of the parameter space were
performed by sampling the posterior probability distribution function with
Monte-Carlo Markov Chains (MCMC). We assumed flat priors. Table \ref{tab:resNoM1}
summarizes the results (``best fit'' values are defined as the median value
of the marginalized PDF) and their corresponding 68\% CL uncertainties after
marginalizing the posterior over all the other parameters. The best fit model
yields a $\chi^2/{\rm dof} = 13.2/11 \simeq 1.20$ which is statistically
reasonable\footnote{A $\chi^2$ distribution with 11 degrees of freedom gives a
  probability of 28\% that the $\chi^2$ value will be greater than $13.2$\ .}.

%.................................
\begin{table}[htbp]
\begin{center}
  \caption{Best-fit model parameters for \lname~using a single lens plane.}\label{tab:resNoM1}
  \begin{tabular}{lc}\hline\hline
    $b_\infty$ [arcsec]      &  $2.54 \pm 0.09$ \\
    $\gamma'$         &  $2.00 \pm 0.03$ \\
    axis ratio $q$    &  $0.869\mypm{0.017}{0.013}$ \\
    $PA_0$  &  $-11.8\mypm{7.0}{8.9}$ \\
    $\gamma_{\rm ext}$ &  $0.067\mypm{0.010}{0.007}$ \\
    $PA_{\rm ext}$ &  $-31.5\mypm{6.9}{4.8}$ \\
    $\zsii$           &  $5.30\mypm{1.03}{1.00}$ \\
    \hline
    $\sigma_{\rm SIE}$ [$\kms$] &  $287.0 \mypm{5.1}{5.3}$ \\
    ``unlensed'' apparent $F814_{s1}$  [mag] & $22.76 \pm 0.02\pm0.10$ \\
    ``unlensed'' absolute $V_{s1}$   [mag] & $-19.79 \pm 0.05\pm0.10$ \\
    ``unlensed'' apparent $F814_{s2}$  [mag] & $27.01 \pm 0.09\pm0.10$ \\
  \hline\end{tabular}
\end{center}
{\small Best fit model parameters and 68.4\% confidence limits. Errors on magnitudes distinguish statistical (first) and systematic from lens light subtraction (second). Angles are in degrees oriented from North to East.}
\end{table}

The results of the best fit model inferred from the conjugation of
bright spots is shown in Fig. \ref{fig:model1} where we used the
pixellized source inversion technique to illustrate the quality of the
fit and the reliability of the conjugation method. Although the
surface brightness of Ring 1 and Ring 2 identified by separate
annuli in the image plane are inverted separately, model predictions
in the image plane are recombined for convenience. The two source
planes $\zsi=0.609$ and $\zsii\simeq 5$ are also shown.

As expected, there is a degeneracy between external shear and
ellipticity of the total mass distribution and the modeling, suggesting that
the major axis of the potential and the external shear differ by
$PA_0 - PA_{\rm ext} = 20\mypm{12}{16}$ deg, that is they are aligned within
$\sim 1.2\sigma$. The orientation of external shear in agreement with the orientation of stars out to $r\lesssim 1\arcsec$ which is about $-36$ deg. The orientation of the internal quadruple (lens ellipticity) and that of stars are misaligned by $\sim 24^\circ$.
 Likewise, the axis ratio
of the light distribution over this radial range is
$0.85 \lesssim b/a\lesssim 0.93$, again consistent with our lens model.

%----------------------------------------
\begin{figure}[htb]
  \centering \includegraphics[width=\hsize]{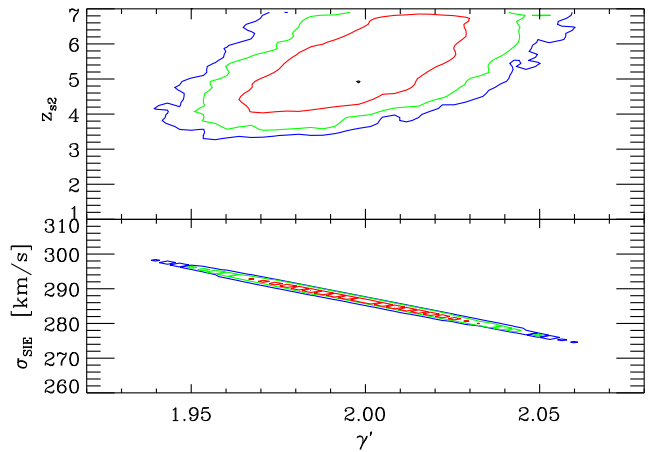}
  \caption{\small {\it Top panel:} 68.3\%, 95.4\% and 99.3\% CL
  contours for model parameters slope of the density profile $\gamma'$
  and source 2 redshift $\zsii$.  \ \ \ {\it Bottom panel:} {\it Idem}
  for the slope $\gamma'$ and the lens equivalent velocity dispersion
  (defined as $186.2 \sqrt{b_\infty q^{1/2}/1\arcsec} \kms$).}\label{fig:cont-prof}
\end{figure}

The lens modeling also puts interesting constraints on the redshift of
source 2: $\zsii=5.3\pm1.0$. The accuracy is relatively low
because of the saturation of the $\dlsdos(\zs)$ curve when $\zs\rightarrow
\infty$. The top panel of Fig. \ref{fig:cont-prof} shows a mild
correlation between $\zsii$ and the slope of the density profile
$\gamma'$. This is expected since the steeper the density profile that
fits the inner ring, the less mass is enclosed between the two rings,
and hence the further away must be the outer source.

\begin{figure}[htb]
  \centering \includegraphics[width=\hsize]{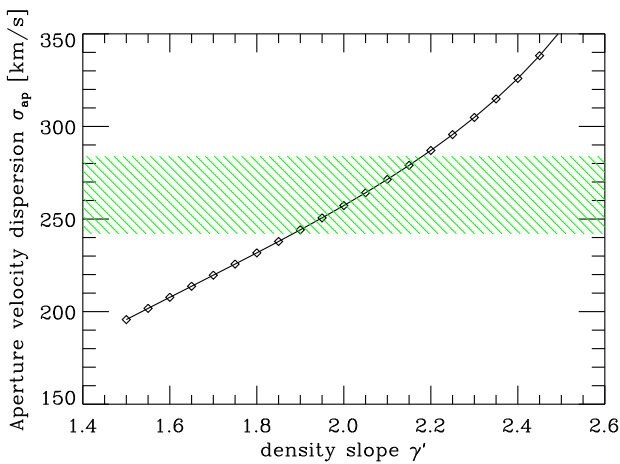}
  \caption{\small Predicted stellar aperture velocity dispersion $\sigma_{\rm ap}$ as it would be measured with SDSS spectroscopic settings as a function of the slope of the density profile. The normalization of density profile is fixed to be consistent with the Einstein radius of Ring 1. The shaded area is the 1$\sigma$ SDSS measurement uncertainty. It shows a remarkable agreement between the double source plane analysis and the coupling of kinematical + source 1 plane data, both favoring nearly isothermal slopes. Note that $\sigma_{\rm ap}$ and $\sigma_{\rm SIE}$ do not need to be identical.} \label{fig:kinem+ring1}
\end{figure}

In spite of the complexity of the azimuthal properties of the lens
potential, our modeling yielded stable and well localized constraints
on the normalization and slope of the radial total density
profile. The lower panel of Fig. \ref{fig:cont-prof} shows the
confidence regions for the slope $\gamma'$ and the equivalent velocity
dispersion $\sigma_{\rm SIE}$. First, we find a total density profile very
close to isothermal with a slope $\gamma'=2.00\pm0.03$. The
corresponding SIE velocity dispersion is $\sigma_{\rm SIE}=287.0\pm 5.2\kms$.
In order to compare these results with SDSS spectroscopy, one needs to
solve the spherical Jeans equation taking into account observational
effects (SDSS fiber aperture, seeing) and the surface density of dynamical
tracers (radial distribution of stars in the lens galaxy) measured in
\S\ref{ssec:lens-pptes}. Here we assume an isotropic pressure tensor.
A general description of the method can be found in \citet{koopmans06b}.
Fig. \ref{fig:kinem+ring1} shows the aperture velocity dispersion that
would  be measured with SDSS spectroscopic fibers when the density profile
is normalized to fit the first ring alone. It shows that slopes close to
isothermal $(\gamma'\simeq2)$ predict velocity dispersions close to SDSS
spectroscopic velocity dispersion,
which gives strong support to our double source plane lensing-only analysis.
Such a similarity is consistent with the results of previous SLACS studies
\citep{treu06,koopmans06}. We note that the accuracy reached on both the
slope and the velocity dispersion based on lensing constraints alone
is better than that afforded by kinematical measurements at the same
redshift, although the two methods are complementary in their
systematic errors and degeneracies \citep[see discussion in \eg][]{treu02}.

%----------------------------------------

%------------------------------------------------------------
\subsection{Budget of mass and light in \lname}\label{ssec:moverl}
The tight constraints on the projected mass profile between the two
Einstein radii can be compared to the light distribution inferred in
\S\ref{ssec:lens-pptes}. In particular, the total projected V band
mass-to-light ratio within the effective radius $\reff\simeq 7.29
\hmkpc$ is $M/L_V=11.54\pm0.51 \hmmolsun$ (corresponding to a total projected
mass $4.90\pm0.13\times 10^{11} \hmmsun$). The logarithmic slope of the
projected enclosed total mass profile is $3-\gamma'=1.00\pm0.03$,
while the slope of the cumulative luminosity profile close to the
effective radius is $\der \log L(<r) /\der \log r=0.62$ with much smaller
uncertainty. Therefore the projected mass-to-light ratio profile increases
with radius as $r^{0.38\pm0.03}$ around $\reff$ with high statistical
significance.

We now compare these values to the stellar mass content in the effective
radius using a  the typical mass-to-light ratio of stellar populations in
massive galaxies at that redshift 
$M_*/L_V\simeq3.14\pm0.32\hmmolsun$ \citep{gavazzi07b} and
$\sim30\%$ intrinsic scatter about this values (due to \eg~age-metallicity
effects) as found in the local Universe \citep{gerhard01,trujillo04}.
This leads to a fraction of projected mass in the form of dark matter 
within the effective radius $f_{\rm DM,2D}(<\reff)\simeq 73\pm9\%$ which
is about twice as high as the average value found by \citet{gavazzi07b}
and \citet{koopmans06} thus making \lname~a particularly dark-matter-rich system.

%======================================================================
%======================================================================
\section{Exploiting the double source plane: beyond the lens mass properties}
\label{sec:beyond}

In this section we address two particular applications afforded by
the double source plane nature of \lname. First, in
\S\ref{ssec:cosmo} we discuss whether this particular system  gives
interesting constraints on cosmological parameters. Then, in
\S\ref{ssec:Ms1}, we present a compound double lens plane mass model and use it to
constrain the total mass of the Ring 1. This provides a new (and
perhaps unique) way to obtain total masses of such compact and faint
objects. Thus, in combination with the magnifying power of the main lens,
this application appears to be a promising way to shed light on the
nature of faint blue compact galaxies \citep[e.g.][]{marshall07}.
In \S~\ref{ssec:gen-cosmo} we discuss the
prospects of doing cosmography with samples of double source plane lenses,
 taking into account the lensing effects of the inner ring on the outer ring.

%------------------------------------------------------------
\subsection{An ideal optical bench for cosmography?} \label{ssec:cosmo}
Can a double source plane lens be used to constrain global cosmological parameters
like $\Omega_{\rm m}$ or $\Omega_\Lambda$? In principle this can be done
because lensing efficiency depends on the ratio of angular diameter
distances to the source $\dos$ and between the lens and the source
$\dls$ as well as the projected surface mass density
$\Sigma(\vec{\theta})$ in the lens plane. In formulae, writing the
lens potential experienced by light rays coming from a source plane as
redshift $\zs$ as:

\begin{eqnarray}\label{eq:defpot}
  \psi(\vec{\theta},\zs) &= &\frac{4 G}{c^2} \frac{\dol\dls}{\dos} \int \der^2 \theta' \Sigma(\vec{\theta}') \ln\vert \vec{\theta}-\vec{\theta}'\vert\;\\
  &\equiv & \psi_0(\vec{\theta}) \frac{\dls}{\dos}\,,
\end{eqnarray}
and considering two images at positions $\vec{\theta}_1$ and
$\vec{\theta}_2$ coming from source planes at redshift $\zsi$ and
$\zsii$, one can measure the ratio of distance ratios $\eta \equiv
(\dlsdos)_{\zsii} / (\dlsdos)_{\zsi}$ directly from the properties of the 
multiple images, given assumptions on the potential $\psi_0(\vec{\theta})$
and its derivatives defining the deflection, convergence and shear at the
positions of the images. 

Applications of this method to clusters of galaxies with several
multiply imaged systems at different source redshifts -- assuming
simple parametric models for the clusters -- seem to favor
$\Omega_{\rm m} <0.5$ cosmologies \citep{golse02a,soucail04}. However,
unknown systematics lurk under the cluster substructure, which can
introduce significant local perturbations of $\psi_0(\vec{\theta})$.
In principle -- at least judging qualitatively from the smoothness of
the isophotes, and the smoothness of galaxy scale Einstein Rings -- one
could hope that massive elliptical galaxies be less prone to this sort
of systematic because source size is large compared to the substructure
angular scale.

\begin{figure}[htb]
  \centering
  \includegraphics[width=\hsize]{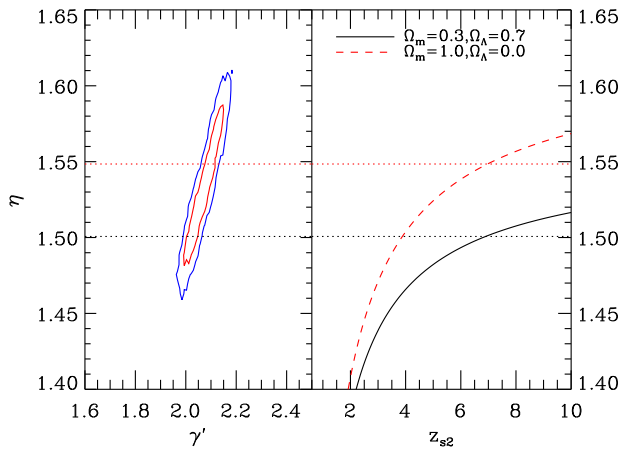}
  \caption{\small {\it Left panel:} Constraints on the logarithmic slope $\gamma'$ and the ratio of distance ratios $\eta$. Contours enclose 68.3\% and 95.4\% of probability. The {\it right panel} shows $\eta(\zsii)$ as a function of $\zsii$ for two flat cosmologies $(\Omega_{\rm m},\Omega_\Lambda)=(0.3,0.7)$ (black) and $(\Omega_{\rm m},\Omega_\Lambda)=(1.,0.)$ (red) which are two sensible ``extreme'' cases. The dotted horizontal lines illustrate the upper limits on $\eta$ for these cosmologies given the assumption $\zsii\le6.9$ (see \S\ref{ssec:lensed}).}
  \label{fig:cosmo-cont}
\end{figure}

In the previous section we constrained $\zsii$ for the given
$\Lambda$CDM concordance cosmology. Here we re-parametrize the problem
using $\eta$ itself as a free parameter to constrain the change in
lensing efficiency between the two source planes. The left panel of
Fig. \ref{fig:cosmo-cont} shows the joint constraints on the two
parameters $\gamma'$ and $\eta$. A first consequence of this more
general parameterization is that, by allowing a broader range of
values for $\eta$ (i.e. allowing more freedom in the cosmological
model), the uncertainties on the slope are significantly increased:
we find $\gamma'=2.07\pm0.06$. Steeper density profiles are now somewhat
compensated by a relatively higher lensing efficiency for the second
source plane. In other words, the tight constraints previously
obtained on the slope of the density profile depend to some extent on
the assumed cosmological model (\ie~assuming $\Lambda$CDM cosmology
led to $\gamma'=2.00\pm0.03$).

% ----------------------------------------

\begin{figure}[htb]
  \centering
  \includegraphics[width=\hsize]{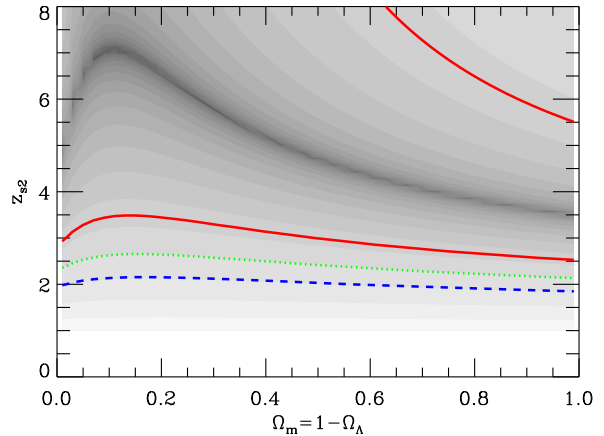}
  \caption{\small 68.3\%, 95.4\% and 99.3\% CL
    contours in the redshift of source 2 and $\Omega_{\rm m}$ parameter space assuming an isothermal density profile. This shows that even using strong priors on the density profile and for a given source redshift, only loose constraints can be inferred on cosmological parameters with a single double source plane system.}
  \label{fig:cosmo-cont2}
\end{figure}

The right panel of Fig.~\ref{fig:cosmo-cont} shows $\eta(\zsii)$ as a
function of the second source redshift for two ``extreme'' flat
cosmologies: $(\Omega_{\rm m},\Omega_\Lambda)=(0.3,0.7)$ and
$(\Omega_{\rm m},\Omega_\Lambda)=(1.,0.)$, intermediate cases lying in
between. This shows that high values $\eta\gtrsim1.57$ are not
consistent with currently favored cosmologies. The upper limit on
$\eta(\zsii=6.9)$ is also shown for these two cases.
This illustrates that very loose constraints can be obtained on cosmological
parameters even if $\zsii$ were known spectroscopically. Likewise, even assuming an
isothermal slope of the density profile as motivated by joint lensing
and dynamical analyses \citep[][]{koopmans06} does not drastically
improve the constraints on $\eta$ and consequently on cosmology as shown
in Fig \ref{fig:cosmo-cont2}, even if $\zsii$ could be measured with
spectroscopic precision.

% ----------------------------------------

However, it is important to point out that the formal $\sim3\%$
relative uncertainty we get on $\eta$ from our lens modeling strategy
based on the identification of conjugate knots underestimates the
potential accuracy of the method. Statistical errors would decrease by
a factor of a few with a full modeling of the surface brightness
distribution in the image plane. Unfortunately, the error budget would
then be limited by additional systematic sources of uncertainty like extra
convergence coming from large scale structures along the line of sight
with estimated standard deviation $\sigma_\kappa \gtrsim 0.02$
\citep{dalal05} or due to a non trivial environment in the main lens plane.
Therefore, we conclude that it is unlikely that any
cosmographic test based on the unique multiple Einstein ring system 
\lname~will provide valuable information on cosmological parameters.
The prospects of using large numbers of double source plane lenses are
investigated in \S\ref{ssec:gen-cosmo}.

%----------------------------------------
\subsection{Source 1, alias Lens 2}\label{ssec:Ms1}

\begin{figure}[htb]
  \centering
  \includegraphics[width=\hsize]{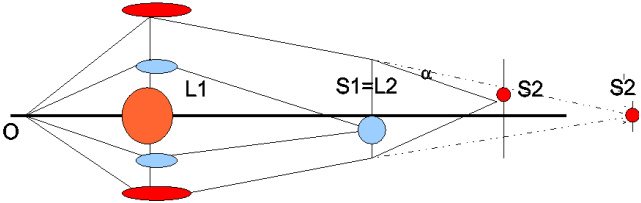}
  \caption{\small Sketch of the lensing optical bench with source 1 acting as a perturbing lens on source 2 which complicates the relation between redshifts, deflection angles and angular distances.}
  \label{fig:bench-sketch}
\end{figure}

Among the massive perturbers along the line of sight to source 2, the
most prominent is probably the mass associated with source 1. Since
the lens modeling predicts that both sources are located very close to
the optical axis (the center of the lens, see lower panels of
Fig. \ref{fig:model1}), the light rays coming from source 2 to the
observer will experience the potential of source 1 before that of the
main lens. Fig. \ref{fig:bench-sketch} illustrates the complexity of
the configuration which adds some extra positive focusing for the second
source plane. For the cosmological applications we mentioned above,
this translates into a small but systematic source of bias.
The bias introduced on the inferred mass profile of the main lens is small,
so that the conclusions presented in \S4 are not significantly altered
except for the estimate of $\zsii$.

% ----------------------------------------

On the bright side, this lens configuration allows us to obtain some
insight on the mass associated with Ring 1 (also identified as 
{\it ``Lens 2''}) provided we now fully take into account the multiple lens
plane nature of such lines of sight \citep[\eg][]{blandford86,SEF92,bartelmann03c}.
This is the purpose of the present section, in which we fix the
$\Lambda$CDM concordance cosmological model for simplicity.

To achieve this goal we have to address the mass properties of the main
lens at the same time as those of the first source 1. We reconsider the
lens model of \S\ref{sec:model}, but add another mass component at
redshift $z_{\rm l2}=\zsi=0.609$ in the form of a singular isothermal
sphere with free equivalent velocity dispersion parameter
$\sigma_{\rm SIE,s1}$ and centered on the position of
source 1.
As in  \S\ref{sec:model}, our uncertainty on the distance to source 2 is
simply parameterized by its redshift $\zsii$ in the context of a
$\Lambda$CDM cosmological model.

In multiple lens-plane theory, the relation between 
the angular position $\vec{\theta}_j$ of a light ray in the $j$-th lens plane and 
the angular position in the $j=1$ image plane is:
\begin{equation}\label{eq:mlp-gen}
  \vec{\theta}_j(\vec{\theta}_1) = \vec{\theta}_1 - \sum_{i=1}^{j-1} \frac{D_{ij}}{D_j} \vec{\hat{\alpha}}(\vec{\theta}_i)\,.
\end{equation}
The last lens plane N can be identified with the source plane such that $\vec{\theta}_N=\vec{\beta}$. In Eq. \ref{eq:mlp-gen}, as compared to \citet{bartelmann03c}, we did not consider the {\it reduced} deflection which introduces an unnecessary extra $D_{i{\rm s}}/D_{\rm s}$ term in the sum. Likewise, the sign convention for the deflection is different than \citet{bartelmann03c}. Therefore for two distinct positions $\vec{\theta}_1$ and $\vec{\theta}_2$ coming from two distinct source plane positions $\vec{\beta}_1$ and $\vec{\beta}_2$ respectively, we can write:
\begin{eqnarray}\label{eq:mlp}
  \vec{\beta}_1 &=& \vec{\theta}_1 - \frac{D_{\rm ls1}}{D_{\rm s1}}\,\vec{\hat{\alpha}}(\vec{\theta}_1) \\
  \vec{\beta}_2 &=& \vec{\theta}_2 - \frac{D_{\rm ls2}}{D_{\rm s2}}\,\vec{\hat{\alpha}}(\vec{\theta}_2) - \frac{D_{\rm s1s2}}{D_{\rm s2}}\,\vec{\hat{\alpha}}_{s1}\left(\vec{\theta}_2- \frac{D_{\rm ls1}}{D_{\rm s1}}\,\vec{\hat{\alpha}}(\vec{\theta}_2) - \vec{\beta}_1\right)\;.
\end{eqnarray}
In these equations, $\hat{\alpha}$ is the deflection produced by the main lensing galaxy (lying in the plane that also defines the image plane) and $\hat{\alpha}_{s1}$ is the perturbing deflection produced by source 1 (lens 2) onto source 2. Note that parameters like the center of source 1 enter
the modeling scheme both as source- and lens-plane parameters.
This is clearly visible in the brackets for the argument of $\hat{\alpha}_{s1}$ that contains $\vec{\beta}_1$,  the position of source 1 in the source plane.

% ----------------------------------------
\begin{figure}[htb]
  \centering
  \includegraphics[width=\hsize]{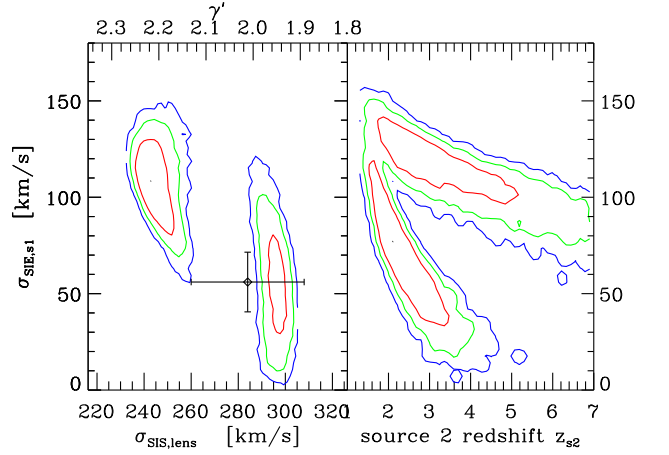}
  \caption{\small {\it Left panel:} contours in parameter space of the velocity dispersion of the main lens $\sigma_{\rm SIE}$ and that of the first source $\sigma_{\rm SIE,s1}$. Given the tight correlation $\sigma_{\rm SIE}\simeq (687 -200.3 \gamma') \kms$ found in \S\ref{ssec:resodel}, the upper abscissa shows the correspondance with slope $\gamma'$. The kinematical SDSS estimate of $\sigma_{v,*}$ and the velocity dispersion of source 1 inferred from the Tully-Fisher relation  \citep{moran07b} are overlaid as a point with error bar. \ \ \ {\it Right panel:} contours in parameter space of the second source redshift $\zsii$ and the velocity dispersion of the first source $\sigma_{\rm SIE,s1}$. The recovered $\zsii$ strongly depends on the mass enclosed in source 1. In both panels confidence levels mark the 68.3, 95.4 and 99.3\% enclosed probability.}
  \label{fig:massS1}
\end{figure}

\begin{figure}[htb]
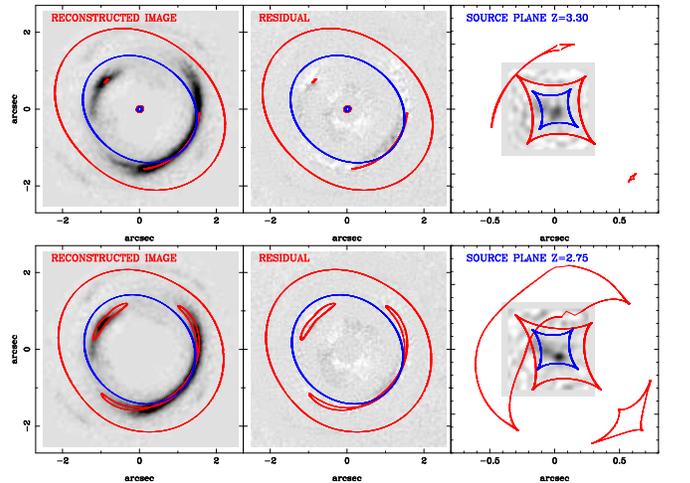

  \centering
  \includegraphics[width=\hsize]{best_fit_lowMs1.eps}
  \includegraphics[width=\hsize]{best_fit_higMs1.eps}
  \caption{\small {\it Top panels:} Best fit family {\it i} model image and source plane reconstructions. From left to right reconstructed image plane, residual (data--model), and source 2 plane at redshift $\zsii=3.30$. {\it Bottom panels:} idem for the best fit family {\it ii} models (with $\zsii=2.75$). Note the complex critical and caustic curves for the $\zsii$ source plane due to the multiple lens plane configuration produced by source 1. For both models the reconstruction is satisfying and produces very few residuals.}
  \label{fig:model_ms1}
\end{figure}

The constraints obtained on the equivalent velocity dispersion parameter
of the main lens $\sigma_{\rm SIE}$ and that of source 1 $\sigma_{\rm SIE,s1}$
are shown in the left panel of Fig. \ref{fig:massS1}. We clearly see two
kinds of solutions: one (family {\it i}) has a high lens velocity
dispersion (and slope $\gamma'\sim1.96$, nearly isothermal) and little
mass in source 1, whereas the other family ({\it ii}) has a lower
main lens velocity dispersion and more mass in source 1. We
measure $(\sigma_{\rm SIE},\sigma_{\rm SIE,s1})=(295\mypm{3.5}{5.0},56\pm{30})
\kms$ for family {\it i} and
$(\sigma_{\rm SIE},\sigma_{\rm SIE,s1})=(247.3\mypm{8.5}{5.7},104\mypm{21}{26})
\kms$ for family {\it ii}. A pixelised source plane inversion for both
of these best fit models is shown in Fig. \ref{fig:model_ms1}. Family
{\it i} models are shown in the top row and family {\it ii} in the
bottom row. Note the very complex systems of caustic and critical lines
produced by this multiple lens plane system. It is difficult to favor
either of these models based on a visual inspection and either region on
the parameter space has about the same statistical weight
(fraction of MCMC samples).

%----------------------------------------

The left panel of Fig. \ref{fig:massS1} also shows the aperture-corrected
SDSS-inferred velocity dispersion of the lens
$\sigma_{v,*}=284\pm24\kms$ which seems to favor family {\it i}
solutions, based on the earlier SLACS results of a general agreement
between stellar velocity dispersion and $\sigma_{\rm SIE}$. In addition,
we can get further external information on the mass of source 1,
by extrapolating the Tully-Fisher relation found by \citet{moran07b}
at $z\sim0.5$ for late-type galaxies. In the field, they found that at
absolute magnitudes of $V\sim -19.7$, the maximum rotation velocity is
$\log(2V_{\rm max})=2.2\pm 0.1$. Assuming $V_{\rm max}\simeq \sqrt{2}
\sigma_{\rm SIE}$, this translates into an estimate  $\sigma_{\rm SIE,s1} \simeq
59\pm 13 \kms$. Another piece of information comes from weak lensing
results at intermediate redshift ($0.2<z<0.4$) by
\citet{hoekstra05a}, who  found that galaxies with magnitude $V-5\log h
\simeq -19$ have virial masses $M_{\rm vir}\simeq
1.50\mypm{0.99}{0.64}\times 10^{11} \hmmsun$ which also corresponds to
$\log(2V_{\rm max})=2.20\pm0.09$, in good agreement with
\citet{moran07b}. These two arguments also seem to favor family {\it i} 
solutions, i.e. those with more mass in the main lens and less in
source 1.

The right panel of Fig. \ref{fig:massS1} shows the important
degeneracy between the redshift of source 2 and the velocity
dispersion of source 1. We can see that the more massive source 1, the
lower $\zsii$ must be. This demonstrates that any cosmographic test
based on multiple source plane lens systems should carefully consider
the mass in the foreground source as a significant perturbation on
light rays coming from the most distant source. Adding a substantial
amount of mass in source 1 significantly changes the inferred
redshift of source 2 either for family {\it i} models which yields
$\zsii=2.6\mypm{1.0}{0.7}$ or family {\it ii} models yielding
$\zsii=3.8\mypm{1.9}{1.5}$\ . Marginalizing over the whole posterior
PDF gives $\zsii=3.1\mypm{2.0}{1.0}$\ .

%.................................
\begin{table}[htbp]
\begin{center}
  \caption{Best-fit model parameters for \lname~using a compound double lens plane.}\label{tab:resM1}
  \begin{tabular}{lccc}\hline\hline
    Parameter           & family {\it i} & family {\it ii} & global\\\hline
    $b_\infty$ [arcsec]  & $2.65 \mypm{0.07}{0.10}$ & $1.91 \mypm{0.07}{0.06}$ & $1.98 \mypm{0.69}{0.11}$\\
    $\gamma'$           & $1.96 \mypm{0.03}{0.02}$ & $2.23 \mypm{0.03}{0.05}$ & $2.18 \mypm{0.07}{0.22}$ \\
    axis ratio $q$      & $0.889\mypm{0.057}{0.016}$ & $0.816\mypm{0.129}{0.027}$ &$0.879\mypm{0.067}{0.083}$ \\
    $PA_0$              & $-15.9\mypm{9.5}{12.2}$ & $-17.9\mypm{9.2}{17.3}$ & $-17.0\mypm{9.3}{15.5}$\\
    $\gamma_{\rm ext}$   & $0.069\mypm{0.016}{0.009}$ & $0.089\mypm{0.026}{0.012}$ & $0.082\mypm{0.026}{0.016}$\\
    $PA_{\rm ext}$       & $-27.6\mypm{6.1}{6.7}$ & $-26.5\mypm{6.2}{6.7}$ &  $-27.0\mypm{6.2}{6.7}$\\
    $\zsii$             & $2.6\mypm{1.0}{0.7}$   & $3.8\mypm{1.9}{1.5}$  &$3.1\mypm{2.0}{1.0}$\\
    $\sigma_{\rm SIE,s1}$
            [$\kms$]    &  $56.6\mypm{30.3}{27.6}$ & $108.9\mypm{18.0}{19.9}$ &  $94.0\mypm{26.7}{46.6}$\\
    \hline
    $\sigma_{\rm SIE}$
               [$\kms$] &  $295\mypm{3}{4}$ & $246\mypm{7}{5}$ & $254 \mypm{43}{11}$ \\
  \hline\end{tabular}
\end{center}
{\small Best fit model parameters and 68.4\% confidence limits. Angles are in degrees oriented from North to East.}
\end{table}

%----------------------------------------
\subsection{Future outlook: cosmography with many double source plane lenses}
\label{ssec:gen-cosmo}

In \S\ref{ssec:cosmo} we explored the possibility of constraining cosmology
with \lname, and came to the conclusion that the errors are too large for
this to be interesting. In \S\ref{ssec:Ms1} we saw that the mass of
the closest source must be taken into account as a perturbation along
the double source plane optical bench. Here we attempt to address the
possibility of using large numbers of such multiple lensing systems to
probe the cosmology. Future space-based missions
like DUNE or JDEM should provide us with tens of thousands of lenses,
among which several tens would be double source plane systems. We also
assume that redshifts will be available, from space- or ground-based
spectroscopic follow-up.

First, we summarize the error budget expected for a typical double
source plane system. As described before, the main quantity of
interest is the ratio of distance ratios parameter
$\eta\equiv (\dlsdos)_2 / (\dlsdos)_1 $, where source 2 is the furthest one.
For simplicity, we assume that the main lens, the first source, and the
second source are perfectly aligned onto the optical axis, resulting
in two complete concentric rings of radius $\theta_1$ and $\theta_2$.
The lens equation for each source plane reads:

\begin{subequations}\label{eq:EqsLens}
\begin{eqnarray}
  \beta_1 &= & \theta_1 - (\dlsdos)_1 \alpha_{\rm tot}(\theta_1) = 0 \,,\label{eq:EqsLensa}\\
  \beta_2 &= &\theta_2 - (\dlsdos)_2 \alpha_{\rm tot}(\theta_2) = 0 \,.\label{eq:EqsLensb}
\end{eqnarray}
\end{subequations}

We consider again the general power-law surface mass distribution of
Eq. \eqref{eq:densprof} producing deflections $\alpha_1$ and $\alpha_2$ on
source 1 and source 2 light rays. For source 2 we must add $\alpha_p$
the {\it small} perturbing deflection\footnote{We assume that the
non-linear coupling between lens planes can be neglected, i.e. the
perturbation of source 1 is small compared to the deflection from the
main lens on source 2 light rays: $\alpha_p\ll \alpha_2
\simeq\theta_2$ .} due to source 1 and experienced by source 2
only. Combining Eq. \eqref{eq:EqsLensa} and Eq. \eqref{eq:EqsLensb} gives:

\begin{equation}\label{eq:eta-comps}
  \eta = \left( \frac{\theta_2}{\theta_1}\right)^{\gamma'-1} \frac{1}{1+\frac{D_{\rm s1s2}}{D_{\rm ls2}}\frac{\alpha_P}{\alpha_2}}\;.
\end{equation}

This equation shows the importance of the perturbation. If one aims at
constraining $\eta$ with interesting accuracy (i.e. error smaller than
0.01), the small perturbing term in the denominator of second part on
the right hand side of Eq. \eqref{eq:eta-comps} should be smaller than 0.01.
Keeping in mind that for lensing potentials close to isothermal, $\alpha \propto
\sigma^2$, and that the typical velocity dispersion of the main lens
is about $\sigma\simeq250\kms$, it is important to control and
correct perturbing potentials with velocity dispersion as small as
$\sigma_{p}=\sigma/10\sim 30 \kms$ for values $D_{\rm s1s2}/D_{\rm ls2}\simeq0.5$.

Next, differentiating Eq. \eqref{eq:eta-comps}, and writing $r\equiv
\theta_2/\theta_1$, one can infer the fractional error on $\eta$:

\begin{align}\label{eq:eta-compsErr}
\left(\frac{\delta_\eta}{\eta}\right)^2 & = &(\gamma'-1)^2  \left(\frac{\delta_r}{r}\right)^2 + (\ln r)^2 \delta_{\gamma'}^2  +  \nonumber\\
 && \frac{4}{\left(1+\frac{D_{\rm ls2}}{D_{\rm s1s2}}\frac{\sigma^2}{\sigma_p^2}\right)^2}
 \left(\frac{\delta_{\sigma_p}}{\sigma_p}\right)^2 \;.
\end{align}

The first contribution is the relative measurement error on the ratio
of Einstein radii, with typical values $0.001 \le \delta_r /r \le
0.03$ for deep space based imaging. The second term captures
our prior uncertainty on the slope of the density profile (for example
\citet{koopmans06} measured $\langle \gamma'\rangle \simeq 2.01$ and
an intrinsic scatter $\delta_{\gamma'}\simeq 0.12$). Finally, the
third term represents our prior knowledge of the mass of the
perturber, which can be based, for example, on the Tully-Fisher relation.
\citet{moran07b} estimated $\delta_{\sigma_p}/\sigma_p\simeq0.25$.
Inserting those values into \eqref{eq:eta-compsErr}, and
assuming a typical value of $r\simeq 1.5$, we find a relative
uncertainty on $\delta_\eta/\eta \simeq 0.06$ for a single system. The
error is dominated by model uncertainties on the slope of the density
profile and the mass in source 1. In the case of \lname, we achieve a
similar accuracy when using the above priors on the slope and the
velocity dispersion of source 1. In the following we shall use a
conservative $\delta_\eta/\eta =0.08$ fiducial value.

Having estimated the accuracy achievable on $\eta$ for a single double
source plane system, we focus on the cosmological meaning of $\eta$ in a
spatially flat universe dominated by Dark Matter ($\Omega_{\rm m}$), and Dark
Energy ($\Omega_{\rm DE}=1-\Omega_{\rm m}$) with equation of state parameter
$w=p_{\rm DE}/\rho_{\rm DE}$. It is worth pointing out that the ratio
of angular diameter distances is independent of the Hubble constant
$H_0$. 

\begin{figure}[htb]
  \centering
  \includegraphics[width=0.48\hsize,height=6cm]{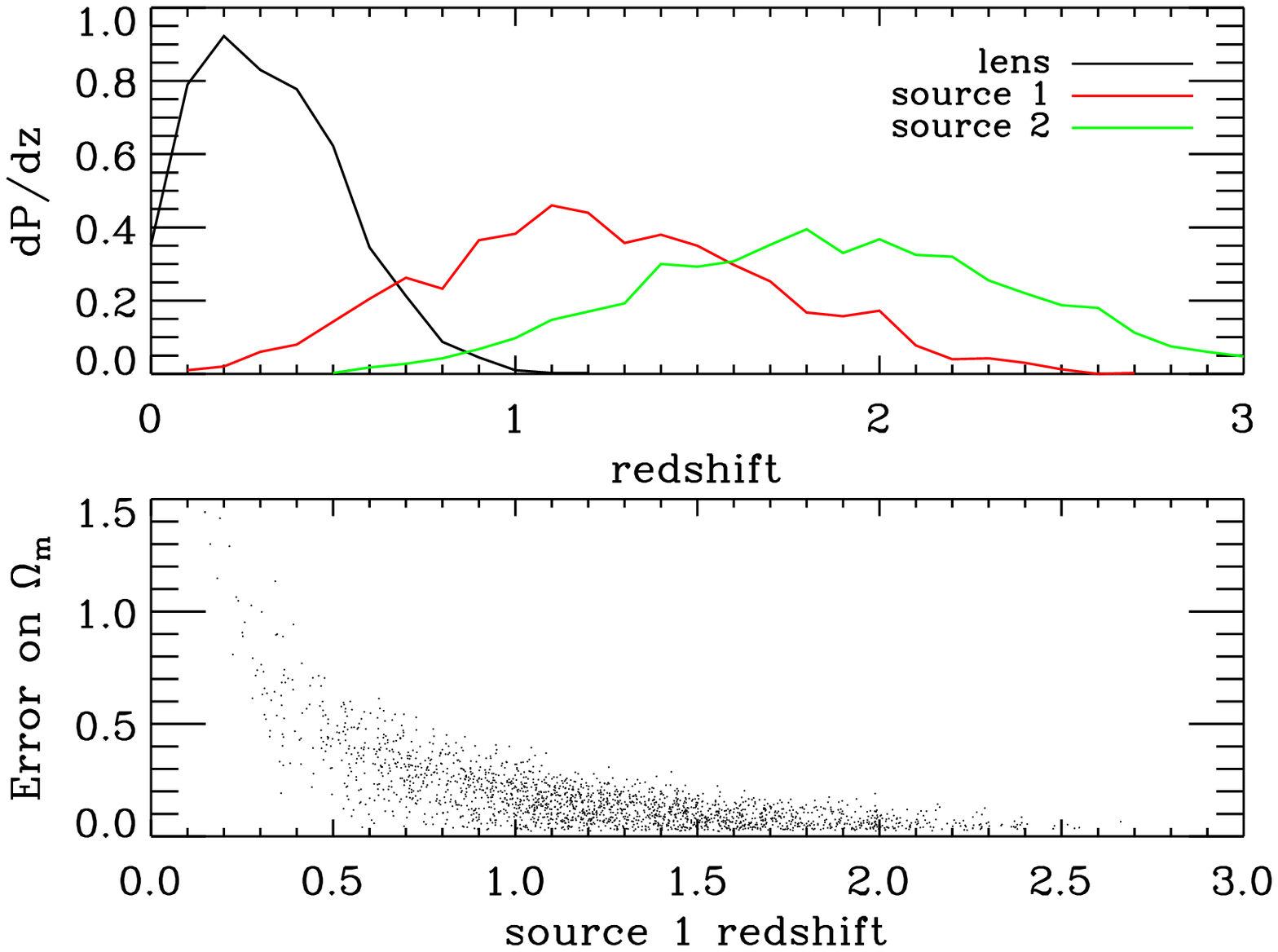}
  \includegraphics[width=0.48\hsize,height=6cm]{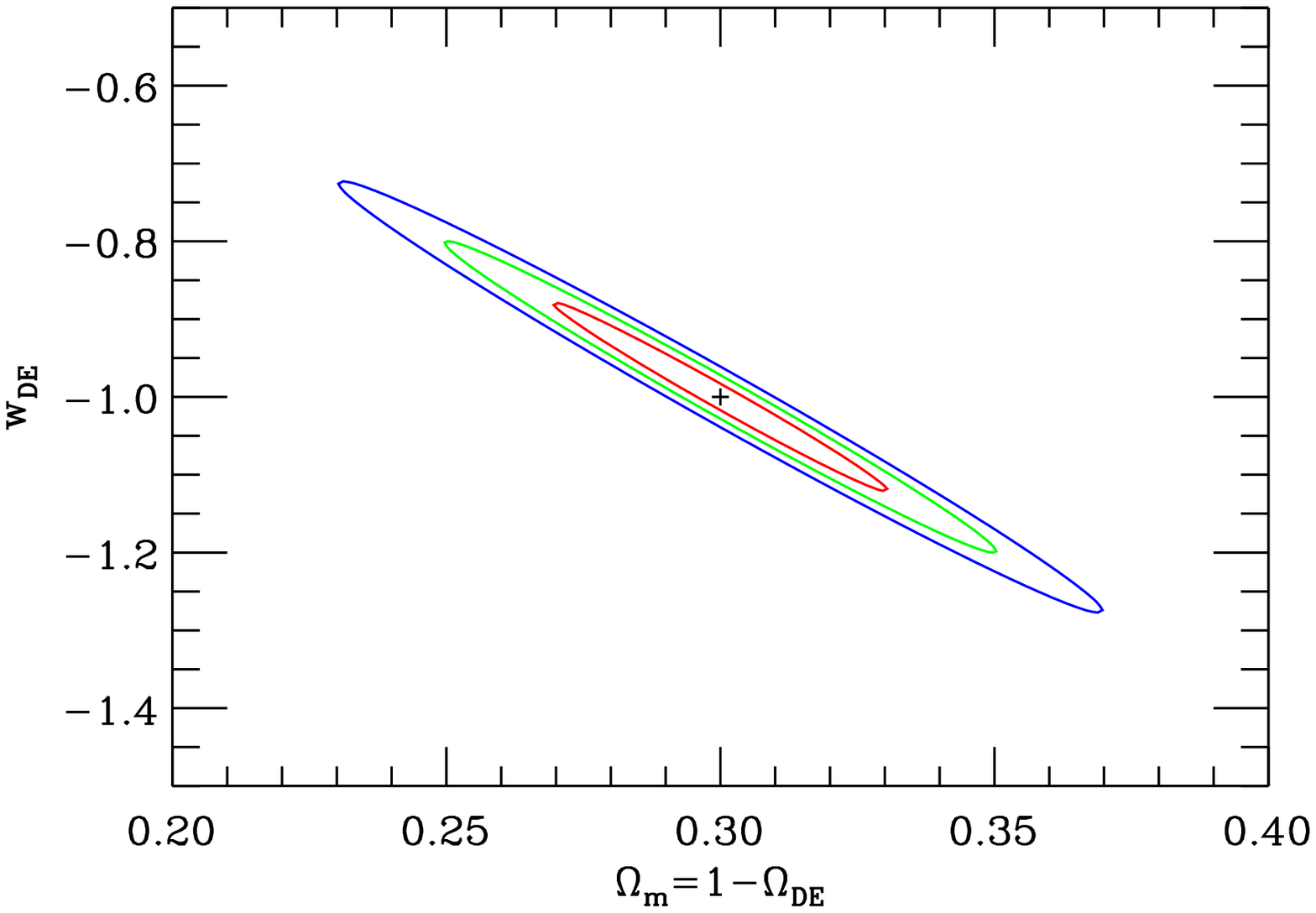}
  \caption{\small {\it Top left panel:} Redshift distribution for the lens, source 1 and source 2 used in the Monte-Carlo simulation of mock double source plane lenses as they could be discovered in future space-based surveys.\ \ {\it Bottom left panel:}  Individual error brought by each system on $\Omega_{\rm m}$ as a function of the redshift of the first source. \ \ \  {\it Right panel:} 68.3, 95.4 and 99.3\% confidence levels contours on the matter density $\Omega_{\rm m}$ and equation of state of dark energy $w=p/\rho$ obtained when combining 50 multiple source plane systems.}
  \label{fig:cosmo-test}
\end{figure}

The constraints on ($\Omega_{\rm m},w$) are shown in the right panel of
Fig. \ref{fig:cosmo-test}. The error contours are obtained using a
Fisher matrix formalism. We assumed a sample of 50 double source plane
lenses, randomly produced using Monte-Carlo simulations. The redshift
distribution of lenses and sources used for the Monte-Carlo simulations
is shown in the upper left panel of Fig. \ref{fig:cosmo-test}.
For the parent population of sources, it is based on recent COSMOS
estimates \citep{leauthaud07}. The equivalent velocity dispersion of the
lenses is assumed to be Gaussian with mean and standard deviation of
$190$ and $60\kms$ respectively. The Einstein radii for the first and second ring
are constrained be greater than $0.7\arcsec$ and $1.0\arcsec$ respectively.

%----------------------------------------

The cosmological parameters $\Omega_{\rm m}$ and $w$ are recovered with
a precision $\pm 0.020$ and $\pm0.080$ respectively. We note that the
sensivity and the orientation of the degeneracy in this set of cosmological
parameters is similar to those obtained with a type Ia supernovae
experiment \citep[see \eg][and references therein]{refregier06}.
The lower left panel of
this figure demonstrates that the systems that contribute the most to
constraining $\Omega_{\rm m}$ are those with source 1 redshift
$\zsi\gtrsim1$ (a similar trend is seen for $w$). Lens redshifts
larger than $\sim 0.5 $ are also more efficient
configurations. This can easily be understood since the higher the redshifts of either lens, first or second source the more sensitive distances are on cosmological parameters. Note that situations with very low redshift lenses but high redshift source 1 and 2 will result in a rapid saturation of the $\dlsdos$ for both source 1 and 2, leading to values $\eta\simeq1$ independant on cosmology. The sensitivity on cosmology is actually essentially driven by the redshift of the primary lens. Therefore we conclude that \lname\, is not an optimal
double source plane lens system for cosmographic purposes.

However the prospects of doing cosmography with future samples of double source plane
lenses are excellent, provided systematic effects are controlled.
The main source of systematic uncertainty that was ignored in the above calculations
is the possibility of a change in the mean density slope as a function of lens
redshift $\langle \gamma'\rangle =f(\zl)$ as present data seem to suggest that
the dynamical properties of early-type galaxies have not changed much since $z\sim1$.
\citet{koopmans06} found $\der \langle \gamma'\rangle / \der z = 0.23\pm0.16$ over
the redshift range [0.08,1.01]. Progress needs to
be made along this line to improve how knowledge of $\langle \gamma'\rangle =f(\zl)$,
but a great advantage of double source plane systems over single ones is that combining
stellar dynamics and lensing constraints from two source planes would be more
efficient at ``self-calibrating'' the method than using single ones.
In addition a thorough lensing analysis aiming at carefully modeling
the surface brightness of lensed structures will certainly help in controlling
any such evolution trend of the density profile \citep[see \eg][]{dye07b}.

%======================================================================
%======================================================================
\section{Summary \& Conclusion}\label{sec:conclu}

In this paper we report the discovery of the first galaxy-scale double
lensing event made of a foreground lens galaxy at redshift $\zl=0.222$,
a first source at redshift $\zsi=0.609$ (Ring 1) and a more distant
source (Ring 2) with unknown redshift, despite an attempt to measure
its redshift with deep optical spectroscopy using LRIS on the Keck I
Telescope. The detection of Ring 2 in a single orbit HST-ACS F814W
filter image, sets an upper limit to its redshift $\zsii < 6.9$.

Modeling the geometry of the lensed features at different source
planes we determine the mass density profile of the lens galaxy which
is found to be close to isothermal. The
best fit lens model predicts a stellar velocity dispersion in very
good agreement with that measured from SDSS spectroscopy.  The model
requires a relatively large amount of dark matter inside the effective
radius $f_{\rm DM,2D}(<\reff)\simeq 73\pm9\%$ (corresponding to a projected
total mass-to-light ratio
$M/L_V=11.54\pm0.13 \hmmolsun$), assuming the stellar mass to light
ratio measured in paper IV. Along with the complex
isophotes of the lens galaxy and the presence of several other (less
luminous) galaxies at similar photometric redshifts, the high dark
matter fraction suggests that the lens may be the central galaxy of a
group scale halo. The high precision of this measurement -- far
superior to that attainable from a single multiply imaged systems --
demonstrates that double source plane lenses are extremely valuable tools
to study the mass profile of galaxies and groups.

In order to constrain the redshift of Ring 2 and assess the
feasibility of determining cosmological parameters using double
source plane lenses, we constructed multiple lens plane mass models.
In that case the lensing effect of Ring 1 on Ring 2 is taken into
account and modeled as a singular isothermal sphere. Although the
extra mass component adds additional uncertainty
to the derived $\zsii$ and cosmological parameters, it provides a
unique way to determine the total mass of the intermediate galaxy
(the inner Ring). Discarding the family {\it ii} range of solutions
(disfavored by kinematics of stars on the main lens galaxy),
the two-lens-plane mass model results can be summarized as follows:

\begin{itemize}

\item The redshift of Ring 2 is found to be
$\zsii=2.6\mypm{1.0}{0.7}$. This is a genuine prediction that can be
tested with the help of deep Hubble images at shorter wavelengths,
using the dropout technique.

\item No interesting constraints on cosmological parameters can be
obtained from the lensing analysis of the system \lname, due to the
unknown redshift of Ring 2, the overall degeneracy of cosmography with
the slope of the mass density profile between the rings, the
degeneracy with the mass of the inner ring, and the suboptimal
combination of lens and source redshifts.

\item The velocity dispersion of Ring 1 is found to be
  $\sigma_{\rm SIE}=56\pm30 \kms$, in good agreement with the value expected
based on the extrapolation of Tully-Fisher relation at this redshift
\citep{moran07b} and on weak-lensing measurements \citep{hoekstra05a}.
Given that lensing cross section increases with
the fourth power of the velocity dispersion, individual lenses in this
mass range are expected to be very rare. In addition, it is observationally more
difficult to identify lensed background structures embedded in foreground
late-type galaxies with small Einstein radius (both for imaging or spectroscopic
lens searches). Thus by exploiting the boost of
the primary lens, double source plane, also seen as double lens plane systems,
may be an effective way
to determine the lensing mass of small distant galaxies, complementing
detailed photometric studies \citep{marshall07}, and kinematic studies
with integral field spectrographs on large ground based telescopes
with adaptive optics.

\end{itemize}

Future planned space missions like JDEM or DUNE are expected to deliver
several tens of thousands of single source plane lenses
\citep{aldering04,marshall05,refregier06} and several tens of double source plane
lens galaxies. Given the great utility of multiple source plane lenses as tools to study distant
galaxies, and the relatively small number of expected systems,
we argue that the necessary effort of spectroscopic follow-up would be easily
affordable and well motivated.

In addition, a relatively large sample of double source plane
galaxy-scale gravitational lenses will be a practical tool for
cosmography. As an example, we calculated the constraints on
$\Omega_{\rm m}$ and the equation of state of Dark Energy $w=p_{\rm
DE}/\rho_{\rm DE}$ that can be obtained from a sample of 50 double source plane
lenses, assuming both source redshifts are known and are realistically
distributed. Spectroscopic follow-up of such systems rings is also required
to control systematic effects such as the change in the mean density profile 
slope as a function of the lens galaxy redshift. 
A careful analysis taking into account the uncertainty on
the mass profile of the main lens and of the perturber shows that
cosmological parameters can be measured with an accuracy of 10\%
comparable to that obtained from the Hubble diagram of Type
Ia supernovae.

%======================================================================
%======================================================================

\acknowledgements
 
This research is supported by NASA through Hubble Space Telescope programs
SNAP-10174, GO-10494, SNAP-10587, GO-10798, GO-10886. TT acknowledges support
from the NSF through CAREER award NSF-0642621, the Sloan Foundation through a
Sloan Research Fellowship.He is also supported by a Packard fellowship.
The work of LAM was carried out at Jet Propulsion Laboratory, California
Institute of Technology under a contract with NASA. L.V.E.K. is supported
in part through an NWO-VIDI program subsidy (project number 639.042.505).
He also acknowledges the continuing support by the European Community's
Sixth Framework Marie Curie Research Training Network Programme,
Contract No. MRTN-CT-2004-505183 ``ANGLES''. PJM acknowledges support
from the Tabasgo foundation in the form of a research fellowship.
PJM is also grateful to Daniel Holz for useful early discussions on double
lenses and their likely frequency. Based on
observations made with the NASA/ESA Hubble Space Telescope, obtained
at the Space Telescope Science Institute, which is operated by the
Association of Universities for Research in Astronomy, Inc., under
NASA contract NAS 5-26555. This project would not have been feasible
without the extensive and accurate database provided by the Digital
Sloan Sky Survey (SDSS). Funding for the creation and distribution of
the SDSS Archive has been provided by the Alfred P. Sloan Foundation,
the Participating Institutions, the National Aeronautics and Space
Administration, the National Science Foundation, the U.S. Department
of Energy, the Japanese Monbukagakusho, and the Max Planck
Society. The SDSS Web site is \url{http://www.sdss.org/}.  The SDSS is
managed by the Astrophysical Research Consortium (ARC) for the
Participating Institutions. The Participating Institutions are The
University of Chicago, Fermilab, the Institute for Advanced Study, the
Japan Participation Group, The Johns Hopkins University, the Korean
Scientist Group, Los Alamos National Laboratory, the
Max-Planck-Institute for Astronomy (MPIA), the Max-Planck-Institute
for Astrophysics (MPA), New Mexico State University, University of
Pittsburgh, University of Portsmouth, Princeton University, the United
States Naval Observatory, and the University of Washington.

%%%%%%%%%%%%%%%%%%%%%%%%%%%%%%%%%%%%%%%%%%%%%%%%%%%%%%%%%%%%%%%%%%%%%%
\appendix
\section{Derivation of the probability of multiple lensing}\label{append:prob}

In this Appendix we estimate the probability of finding a double lens
in a sample of lenses like SLACS. The first ingredient is the surface
density on the sky of potential lens galaxies, given by:

\begin{equation}\label{eq:proba0}
   N_{\rm gal} = \int \der \sigma \int \der \zl\, \frac{ \der n_l}{\der \sigma } p(\zl )\frac{\der V}{\der \zl }  \,,
\end{equation}

with $\frac{\der V}{\der z_l}$ the comoving volume per unit solid
angle and redshift, and $\frac{\der n_l}{\der \sigma}$ the velocity
dispersion function. For simplicity, we assume here that the shape of
the velocity dispersion function does not evolve with redshift, but
only in normalization as described by the $p(\zl)$ function. In
practice, we consider the velocity dispersion function $\frac{\der
n}{\der \sigma}$ measured by \citet{sheth03} at $z\sim0.1$, which is
of the form:

\begin{equation}\label{eq:sigvdf}
\frac{\der n_l}{\der \sigma} = \phi_* \left(\frac{\sigma}{\sigma_*}\right)^\alpha \frac{\beta}{\sigma \Gamma[\alpha/\beta]} \exp\left[ -(\sigma/\sigma_*)^\beta\right].
\end{equation}
with $\phi_*=0.0020\pm0.0001\, {\rm \h^3\,Mpc}^{-3}$,
$\sigma_*=88.8\pm 17.7 \kms$, $\alpha=6.5\pm1.0$ and
$\beta=1.93\pm0.22$\ . 

The number density of foreground galaxies producing a single strong lensing
event on a source population s1 can be written as:
\begin{equation}\label{eq:proba1}
   N_{\rm s1} = \int \der \zsi \int \der \sigma \int \der z_l \frac{\der
   V}{\der z_l}\, \frac{ \der n_l}{\der \sigma} p(\zl )\frac{ \der
   N_{s1}}{\der \zsi } X(\sigma,\zl, \zsi)\,,
\end{equation}
following \citet{marshall05}. In this equation, $X(\sigma,\zl,\zsi)$
is the cross-section for lensing. In most cases of scale-free
gravitational lenses $X(\sigma,\zl,\zsi)$ can be separated such that
$X(\sigma,\zl,\zsi) = \sigma^{2\nu} g(\zl,\zsi)$, $\sigma^\nu$ giving
the overall strength of the lens. For the particular case of a
singular isothermal sphere that we shall consider, $\nu=2$ and
$g\propto (D_{\rm ls1}/D_{\rm os1})^2 \Theta(\zsi-\zl)$, with $\Theta(x)$
the Heaviside step function.

If the lensing cross-section for a second population of sources s2
does not depend on the presence or properties of an already lensed
population s1 galaxy, we can write:
\begin{equation}\label{eq:proba2}
N_{\rm s1,\ s2} = \int \der \zsii  \int \der \zsi \int \der \sigma  \int \der z_l \frac{\der V}{\der z}\, \frac{ \der n_l}{\der \sigma } p(\zl )\frac{ \der N_{s1}}{\der \zsi }\frac{ \der N_{s2}}{\der \zsii }  X_1(\sigma,\zl, \zsi ) X_2(\sigma,\zl,\zsii)\,.
\end{equation}
Combining Eqq. \eqref{eq:proba0}, \eqref{eq:proba1}, \eqref{eq:proba2}
and taking advantage of the separability of the dependency on $\sigma$
and on redshifts, the ratio of the probability that a galaxy lenses a
source at $\zsii$ given that it is already lensing a source at $\zsi$ over
the probability for a galaxy to lens a source at $\zsii$ is given by:
\begin{eqnarray}
 \frac{ P({\rm lens\ s2}\,\vert\,{\rm  lens\  s1})}{ P({\rm lens\  s2})} = &\frac{ P({\rm lens\ s2,\ lens\  s1})}{ P({\rm lens\  s2})\,P({\rm lens\  s1}) } = \frac{ N_{\rm s1,s2} N_{\rm gal}}{ N_{\rm s2}\,N_{\rm s2} }\label{eq:proba3}\\ 
 =&
\frac{ \left[\int \der \sigma \frac{\der n_l}{\der \sigma}\right]\times\left[\int \der \sigma \frac{\der n_l}{\der \sigma} \sigma^8 \right]}{ \left[ \int \der \sigma \frac{\der n_l}{\der \sigma} \sigma^4\right]^2 }
 \frac{\left[\int \der V(z_l) p(\zl )\right]\times \left[\int \der \zsi \frac{ \der N_{s1}}{\der \zsi }\int \der \zsii \frac{ \der N_{s2}}{\der \zsii }\int \der V(z_l) p(\zl ) g(\zl,\zsi) g(\zl,\zsii) \right] }{ \left[\int \der \zsi \frac{ \der N_{s1}}{\der \zsi } \int \der V(z_l) p(\zl ) g(\zl,\zsi)  \right]\times \left[ \int \der \zsii \frac{ \der N_{s2}}{\der \zsii }\int \der V(z_l) p(\zl ) g(\zl,\zsii) \right] }\,\label{eq:proba4}\\
\equiv & \Sigma \, \times\, \zeta\,.
\end{eqnarray}

where $\der V(z_l)$ indicates $\frac{\der V}{\der z_l}\der z_l$.  The
first term $\Sigma$ in Eq. \eqref{eq:proba4} describes the strong
dependency of the lensing cross section on velocity dispersion. As
expected because lensing favors high $\sigma$ systems, using the
velocity dispersion function from \citet{sheth03}, we estimate it to
be larger than unity, of order $\Sigma=\Gamma((8+\alpha)/\beta)
\Gamma(\alpha/\beta)/\Gamma((4+\alpha)/\beta)^2\simeq 2.44$. 

The second term $\zeta$ contains volume and lensing efficiency
$g(\zl,\zs)$ effects that depend on the redshifts of the lens and the
sources. By defining a lensing efficiency averaged over a given
population of sources,
\begin{equation}
G_i(\zl) = \int \der {\zs}_i \frac{\der N_i}{\der {\zs}_i} g(\zl,{\zs}_i)\,,
\end{equation}
we can simplify the second term in Eq. \eqref{eq:proba4} and write it as:
\begin{equation}
\zeta=\frac{ \left[ \int \der V (z_l) p(\zl) \right] \times \left[ \int \der V (z_l) p(\zl)G_1(\zl) G_2(\zl)\right] }{ \left[ \int \der V (z_l) p(\zl)G_1(\zl) \right] \times \left[ \der V (z_l) p(\zl) G_2(\zl) \right] }\,.
\end{equation}

To obtain a quantitative estimate of the probability of double
lensing, let us consider some specific examples. The most important
quantity in the definition of $\zeta$ is the comoving redshift
distribution of deflectors $p(\zl)$. If all of them were confined in a
single lens plane such that $p(\zl)=\delta(\zl-{\zl}_0)$, then
$\zeta=1$ and most of the change in probability comes from selection
effects captured by $\Sigma$. If, instead, deflectors are broadly
distributed over a range of redshifts, the ratio can be significantly
higher than one because the probability of single lensing would be low
for high redshift deflectors whereas the fact that a deflector is
already lensing a source at $\zsi$ favors lenses in a low redshift
range, more suitable for lensing a source at $\zsii$. 

\begin{figure}[htb]
  \centering
  \includegraphics[width=0.5\hsize]{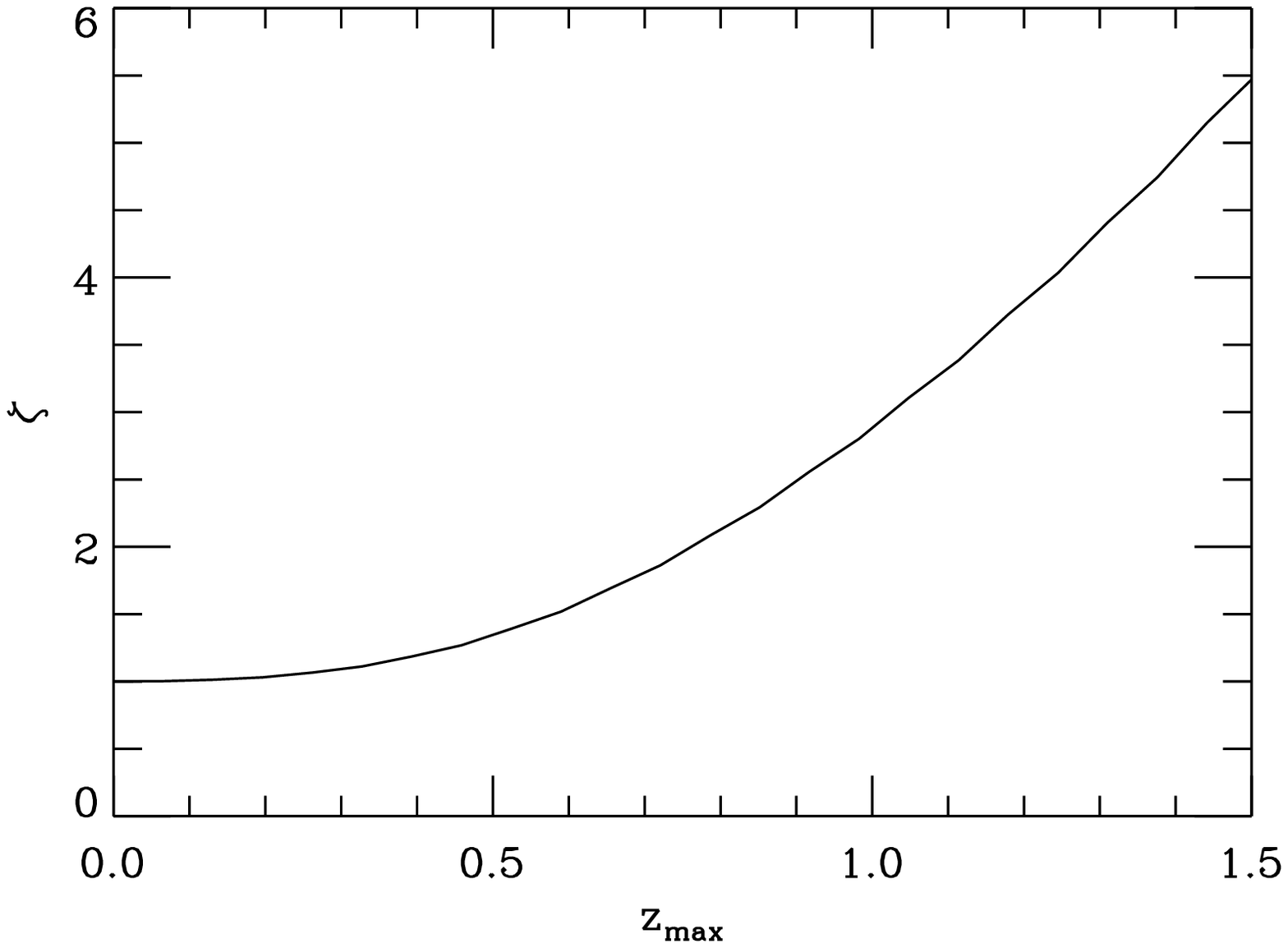}
  \caption{\small Evolution of the $\zeta$ term for multiple lensing probability boost $P({\rm lens\ s2}\,\vert\,{\rm  lens\  s1})/ P({\rm lens\  s2})$ as a function of the limiting redshift $z_{\rm max}$ of the lens distribution.}
  \label{fig:zeta}
\end{figure}

We illustrate this volume effect by assuming that the comoving density
of deflectors is constant out to a redshift $z_{\rm max}$ and then it
drops to zero, that is $p(\zl)=\Theta(z_{\rm max}-\zl)$.
 We also assume the redshift distribution of background sources is of the form
$\der N/\der z \propto {\rm e}^{-z/z_0} (z/z_0)^{a-1}$. Population s2 galaxies
follow the redshift distribution of faint background sources presented in
\citet{gavazzi07b} and having $z_0=0.345$ and $a=3.89$.
We use a redshift distribution for the s1
population that peaks around redshift $0.5$, in agreement with the
properties of spectroscopically discovered SLACS lenses
\citep[see][]{bolton08}. This corresponds to $z_0\simeq 0.07$ and
$a\simeq 7$. Note that
the detailed shape of the redshift distribution for either s1 or s2
galaxies does not change the trends significantly, so this
approximation is sufficient for our purposes. Fig. \ref{fig:zeta}
shows the evolution of $\zeta$ as a function of the limiting redshift
$z_{\rm max}$. We see that, out to reasonable values $z_{\rm
max}\simeq 0.5$, $\zeta$ does not depart much from unity.

To obtain a numerical value to be compared with our SLACS sample, we
consider a population of deflectors constant out to redshift
unity. The gain in probability is $ P({\rm lens\ s2}\,\vert\,{\rm
lens\ s1})/ P({\rm lens\ s2}) \simeq 2.4 - 5$. In other words, if
one elliptical galaxy at $z\lesssim 0.8$ in about 200 is strongly lensing a
faint background source, a strong lens in approximately 40--80 is a
double lens. This is consistent with the observations.

%----------------------------------------

%======================================================================
%=====================================================================
%\bibliographystyle{apj}
\bibliographystyle{aa}
\bibliography{references}

\begin{thebibliography}{58}
\expandafter\ifx\csname natexlab\endcsname\relax\def\natexlab#1{#1}\fi

\bibitem[{{Aldering} \& the SNAP~collaboration(2004)}]{aldering04}
{Aldering}, G. \& the SNAP~collaboration. 2004, astro-ph/0405232

\bibitem[{{Barnab{\`e}} \& {Koopmans}(2007)}]{barnabe07}
{Barnab{\`e}}, M. \& {Koopmans}, L.~V.~E. 2007, \apj, 666, 726

\bibitem[{{Bartelmann}(2003)}]{bartelmann03c}
{Bartelmann}, M. 2003, astro-ph/0304162

\bibitem[{{Bell} {et~al.}(2006){Bell}, {Naab}, {McIntosh}, {Somerville},
  {Caldwell}, {Barden}, {Wolf}, {Rix}, {Beckwith}, {Borch}, {H{\"a}ussler},
  {Heymans}, {Jahnke}, {Jogee}, {Koposov}, {Meisenheimer}, {Peng}, {Sanchez},
  \& {Wisotzki}}]{bell06a}
{Bell}, E.~F., {Naab}, T., {McIntosh}, D.~H., {et~al.} 2006, \apj, 640, 241

\bibitem[{{Blandford} \& {Narayan}(1986)}]{blandford86}
{Blandford}, R. \& {Narayan}, R. 1986, \apj, 310, 568

\bibitem[{{Bolton} {et~al.}(2005){Bolton}, {Burles}, {Koopmans}, {Treu}, \&
  {Moustakas}}]{bolton05a}
{Bolton}, A.~S., {Burles}, S., {Koopmans}, L.~V.~E., {Treu}, T., \&
  {Moustakas}, L.~A. 2005, \apjl, 624, L21

\bibitem[{{Bolton} {et~al.}(2006{\natexlab{a}}){Bolton}, {Burles}, {Koopmans},
  {Treu}, \& {Moustakas}}]{bolton06}
{Bolton}, A.~S., {Burles}, S., {Koopmans}, L.~V.~E., {Treu}, T., \&
  {Moustakas}, L.~A. 2006{\natexlab{a}}, \apj, 638, 703

\bibitem[{{Bolton} {et~al.}(2004){Bolton}, {Burles}, {Schlegel}, {Eisenstein},
  \& {Brinkmann}}]{bolton04}
{Bolton}, A.~S., {Burles}, S., {Schlegel}, D.~J., {Eisenstein}, D.~J., \&
  {Brinkmann}, J. 2004, \aj, 127, 1860

\bibitem[{{Bolton} {et~al.}(2007){Bolton}, {Burles}, {Treu}, {Koopmans}, \&
  {Moustakas}}]{bolton07a}
{Bolton}, A.~S., {Burles}, S., {Treu}, T., {Koopmans}, L.~V.~E., \&
  {Moustakas}, L.~A. 2007, \apjl, 665, L105

\bibitem[{{Bolton} {et~al.}(2008){Bolton}, {Burles}, {Treu}, {Koopmans},
  {Moustakas}, {Gavazzi}, \& {Wayth}}]{bolton08}
{Bolton}, A.~S., {Burles}, S., {Treu}, T., {et~al.} 2008, in preparation

\bibitem[{{Bolton} {et~al.}(2006{\natexlab{b}}){Bolton}, {Moustakas}, {Stern},
  {Burles}, {Dey}, \& {Spinrad}}]{bolton06b}
{Bolton}, A.~S., {Moustakas}, L.~A., {Stern}, D., {et~al.} 2006{\natexlab{b}},
  \apjl, 646, L45

\bibitem[{{Brainerd} {et~al.}(1996){Brainerd}, {Blandford}, \&
  {Smail}}]{brainerd96}
{Brainerd}, T.~G., {Blandford}, R.~D., \& {Smail}, I. 1996, \apj, 466, 623

\bibitem[{{Brewer} \& {Lewis}(2006)}]{brewer06}
{Brewer}, B.~J. \& {Lewis}, G.~F. 2006, \apj, 637, 608

\bibitem[{{Cabanac} {et~al.}(2007){Cabanac}, {Alard}, {Dantel-Fort}, {Fort},
  {Gavazzi}, {Gomez}, {Kneib}, {Le F{\`e}vre}, {Mellier}, {Pello}, {Soucail},
  {Sygnet}, \& {Valls-Gabaud}}]{cabanac07}
{Cabanac}, R.~A., {Alard}, C., {Dantel-Fort}, M., {et~al.} 2007, \aap, 461, 813

\bibitem[{{Casertano} {et~al.}(2000){Casertano}, {de Mello}, {Dickinson},
  {Ferguson}, {Fruchter}, {Gonzalez-Lopezlira}, {Heyer}, {Hook}, {Levay},
  {Lucas}, {Mack}, {Makidon}, {Mutchler}, {Smith}, {Stiavelli}, {Wiggs}, \&
  {Williams}}]{casertano00}
{Casertano}, S., {de Mello}, D., {Dickinson}, M., {et~al.} 2000, \aj, 120, 2747

\bibitem[{{Conroy} {et~al.}(2007){Conroy}, {Prada}, {Newman}, {Croton}, {Coil},
  {Conselice}, {Cooper}, {Davis}, {Faber}, {Gerke}, {Guhathakurta}, {Klypin},
  {Koo}, \& {Yan}}]{conroy07}
{Conroy}, C., {Prada}, F., {Newman}, J.~A., {et~al.} 2007, \apj, 654, 153

\bibitem[{{Dalal} {et~al.}(2005){Dalal}, {Hennawi}, \& {Bode}}]{dalal05}
{Dalal}, N., {Hennawi}, J.~F., \& {Bode}, P. 2005, \apj, 622, 99

\bibitem[{{Dye} {et~al.}(2007){Dye}, {Smail}, {Swinbank}, {Ebeling}, \&
  {Edge}}]{dye07}
{Dye}, S., {Smail}, I., {Swinbank}, A.~M., {Ebeling}, H., \& {Edge}, A.~C.
  2007, \mnras, 379, 308

\bibitem[{{Dye} \& {Warren}(2007)}]{dye07b}
{Dye}, S. \& {Warren}, S. 2007, astro-ph/0708.0787, 708

\bibitem[{{Gavazzi} {et~al.}(2003){Gavazzi}, {Fort}, {Mellier}, {Pell{\' o}},
  \& {Dantel-Fort}}]{gavazzi03}
{Gavazzi}, R., {Fort}, B., {Mellier}, Y., {Pell{\' o}}, R., \& {Dantel-Fort},
  M. 2003, \aap, 403, 11

\bibitem[{{Gavazzi} {et~al.}(2007){Gavazzi}, {Treu}, {Rhodes}, {Koopmans},
  {Bolton}, {Burles}, {Massey}, \& {Moustakas}}]{gavazzi07b}
{Gavazzi}, R., {Treu}, T., {Rhodes}, J.~D., {et~al.} 2007, \apj, 667, 176

\bibitem[{{Gerhard} {et~al.}(2001){Gerhard}, {Kronawitter}, {Saglia}, \&
  {Bender}}]{gerhard01}
{Gerhard}, O., {Kronawitter}, A., {Saglia}, R.~P., \& {Bender}, R. 2001, \aj,
  121, 1936

\bibitem[{{Golse} {et~al.}(2002){Golse}, {Kneib}, \& {Soucail}}]{golse02a}
{Golse}, G., {Kneib}, J.-P., \& {Soucail}, G. 2002, \aap, 387, 788

\bibitem[{{Hoekstra} {et~al.}(2005){Hoekstra}, {Hsieh}, {Yee}, {Lin}, \&
  {Gladders}}]{hoekstra05a}
{Hoekstra}, H., {Hsieh}, B.~C., {Yee}, H.~K.~C., {Lin}, H., \& {Gladders},
  M.~D. 2005, \apj, 635, 73

\bibitem[{{Hoekstra} {et~al.}(2004){Hoekstra}, {Yee}, \&
  {Gladders}}]{hoekstra04}
{Hoekstra}, H., {Yee}, H.~K.~C., \& {Gladders}, M.~D. 2004, \apj, 606, 67

\bibitem[{{Kochanek} {et~al.}(1999){Kochanek}, {Falco}, {Impey}, {Lehar},
  {McLeod}, \& {Rix}}]{kochanek99}
{Kochanek}, C.~S., {Falco}, E.~E., {Impey}, C.~D., {et~al.} 1999, in American
  Institute of Physics Conference Series, Vol. 470, After the Dark Ages: When
  Galaxies were Young (the Universe at 2<Z< 5), ed. S.~{Holt} \& E.~{Smith},
  163--+

\bibitem[{{Kochanek} \& {Narayan}(1992)}]{kochanek92}
{Kochanek}, C.~S. \& {Narayan}, R. 1992, \apj, 401, 461

\bibitem[{{Koopmans}(2005)}]{Koopmans05}
{Koopmans}, L.~V.~E. 2005, \mnras, 363, 1136

\bibitem[{{Koopmans}(2006)}]{koopmans06b}
{Koopmans}, L.~V.~E. 2006, in Engineering and Science, Vol.~20, EAS
  Publications Series, ed. G.~A. {Mamon}, F.~{Combes}, C.~{Deffayet}, \&
  B.~{Fort}, 161--166

\bibitem[{{Koopmans} {et~al.}(2006){Koopmans}, {Treu}, {Bolton}, {Burles}, \&
  {Moustakas}}]{koopmans06}
{Koopmans}, L.~V.~E., {Treu}, T., {Bolton}, A.~S., {Burles}, S., \&
  {Moustakas}, L.~A. 2006, \apj, 649, 599

\bibitem[{{Leauthaud} {et~al.}(2007){Leauthaud}, {Massey}, {Kneib}, {Rhodes},
  {Johnston}, {Capak}, {Heymans}, {Ellis}, {Koekemoer}, {Le F{\`e}vre},
  {Mellier}, {R{\'e}fr{\'e}gier}, {Robin}, {Scoville}, {Tasca}, {Taylor}, \&
  {Van Waerbeke}}]{leauthaud07}
{Leauthaud}, A., {Massey}, R., {Kneib}, J.-P., {et~al.} 2007, \apjs, 172, 219

\bibitem[{{Mandelbaum} {et~al.}(2006){Mandelbaum}, {Seljak}, {Kauffmann},
  {Hirata}, \& {Brinkmann}}]{mandelbaum06}
{Mandelbaum}, R., {Seljak}, U., {Kauffmann}, G., {Hirata}, C.~M., \&
  {Brinkmann}, J. 2006, \mnras, 368, 715

\bibitem[{{Marshall} {et~al.}(2005){Marshall}, {Blandford}, \&
  {Sako}}]{marshall05}
{Marshall}, P., {Blandford}, R., \& {Sako}, M. 2005, New Astronomy Review, 49,
  387

\bibitem[{{Marshall} {et~al.}(2007){Marshall}, {Treu}, {Melbourne}, {Gavazzi},
  {Bundy}, {Ammons}, {Bolton}, {Burles}, {Larkin}, {Le Mignant}, {Koo},
  {Koopmans}, {Max}, {Moustakas}, {Steinbring}, \& {Wright}}]{marshall07}
{Marshall}, P.~J., {Treu}, T., {Melbourne}, J., {et~al.} 2007, ApJ in press,
  astro-ph/0710.0637, 710

\bibitem[{{Moran} {et~al.}(2007){Moran}, {Loh}, {Ellis}, {Treu}, {Bundy}, \&
  {MacArthur}}]{moran07b}
{Moran}, S.~M., {Loh}, B.~L., {Ellis}, R.~S., {et~al.} 2007, \apj, 665, 1067

\bibitem[{{Moustakas} {et~al.}(2007){Moustakas}, {Marshall}, {Newman}, {Coil},
  {Cooper}, {Davis}, {Fassnacht}, {Guhathakurta}, {Hopkins}, {Koekemoer},
  {Konidaris}, {Lotz}, \& {Willmer}}]{moustakas07}
{Moustakas}, L.~A., {Marshall}, P., {Newman}, J.~A., {et~al.} 2007, \apjl, 660,
  L31

\bibitem[{{Myers} {et~al.}(2003){Myers}, {Jackson}, {Browne}, {de Bruyn},
  {Pearson}, {Readhead}, {Wilkinson}, {Biggs}, {Blandford}, {Fassnacht},
  {Koopmans}, {Marlow}, {McKean}, {Norbury}, {Phillips}, {Rusin}, {Shepherd},
  \& {Sykes}}]{myers03b}
{Myers}, S.~T., {Jackson}, N.~J., {Browne}, I.~W.~A., {et~al.} 2003, \mnras,
  341, 1

\bibitem[{{Oyaizu} {et~al.}(2007){Oyaizu}, {Lima}, {Cunha}, {Lin}, {Frieman},
  \& {Sheldon}}]{oyaizu07}
{Oyaizu}, H., {Lima}, M., {Cunha}, C.~E., {et~al.} 2007, astro-ph/0708.0030,
  708

\bibitem[{{Peng} {et~al.}(2002){Peng}, {Ho}, {Impey}, \& {Rix}}]{peng02}
{Peng}, C.~Y., {Ho}, L.~C., {Impey}, C.~D., \& {Rix}, H. 2002, \aj, 124, 266

\bibitem[{{Prada} {et~al.}(2003){Prada}, {Vitvitska}, {Klypin}, {Holtzman},
  {Schlegel}, {Grebel}, {Rix}, {Brinkmann}, {McKay}, \& {Csabai}}]{prada03}
{Prada}, F., {Vitvitska}, M., {Klypin}, A., {et~al.} 2003, \apj, 598, 260

\bibitem[{{Ratnatunga} {et~al.}(1999){Ratnatunga}, {Griffiths}, \&
  {Ostrander}}]{ratnatunga99}
{Ratnatunga}, K.~U., {Griffiths}, R.~E., \& {Ostrander}, E.~J. 1999, \aj, 117,
  2010

\bibitem[{{R{\'e}fr{\'e}gier} {et~al.}(2006){R{\'e}fr{\'e}gier}, {Boulade},
  {Mellier}, {Milliard}, {Pain}, {Michaud}, {Safa}, {Amara}, {Astier},
  {Barrelet}, {Bertin}, {Boulade}, {Cara}, {Claret}, {Georges}, {Grange},
  {Guy}, {Koeck}, {Kroely}, {Magneville}, {Palanque-Delabrouille}, {Regnault},
  {Smadja}, {Schimd}, \& {Sun}}]{refregier06}
{R{\'e}fr{\'e}gier}, A., {Boulade}, O., {Mellier}, Y., {et~al.} 2006, in
  Presented at the Society of Photo-Optical Instrumentation Engineers (SPIE)
  Conference, Vol. 6265, Space Telescopes and Instrumentation I: Optical,
  Infrared, and Millimeter. Edited by Mather, John C.; MacEwen, Howard A.; de
  Graauw, Mattheus W. M.. Proceedings of the SPIE, Volume 6265, pp. 62651Y
  (2006).

\bibitem[{{Rubin} {et~al.}(1980){Rubin}, {Peterson}, \& {Ford}}]{rubin80}
{Rubin}, V.~C., {Peterson}, C.~J., \& {Ford}, Jr., W.~K. 1980, \apj, 239, 50

\bibitem[{{Schlegel} {et~al.}(1998){Schlegel}, {Finkbeiner}, \&
  {Davis}}]{schlegel98}
{Schlegel}, D.~J., {Finkbeiner}, D.~P., \& {Davis}, M. 1998, \apj, 500, 525

\bibitem[{{Schneider} {et~al.}(1992){Schneider}, {Ehlers}, \& {Falco}}]{SEF92}
{Schneider}, P., {Ehlers}, J., \& {Falco}, E.~E. 1992, {Gravitational Lenses}
  (Springer-Verlag Berlin Heidelberg New York)

\bibitem[{{Sheldon} {et~al.}(2004){Sheldon}, {Johnston}, {Frieman}, {Scranton},
  {McKay}, {Connolly}, {Budav{\' a}ri}, {Zehavi}, {Bahcall}, {Brinkmann}, \&
  {Fukugita}}]{sheldon04}
{Sheldon}, E.~S., {Johnston}, D.~E., {Frieman}, J.~A., {et~al.} 2004, \aj, 127,
  2544

\bibitem[{{Sheth} {et~al.}(2003){Sheth}, {Bernardi}, {Schechter}, {Burles},
  {Eisenstein}, {Finkbeiner}, {Frieman}, {Lupton}, {Schlegel}, {Subbarao},
  {Shimasaku}, {Bahcall}, {Brinkmann}, \& {Ivezi{\'c}}}]{sheth03}
{Sheth}, R.~K., {Bernardi}, M., {Schechter}, P.~L., {et~al.} 2003, \apj, 594,
  225

\bibitem[{{Soucail} {et~al.}(2004){Soucail}, {Kneib}, \& {Golse}}]{soucail04}
{Soucail}, G., {Kneib}, J.-P., \& {Golse}, G. 2004, \aap, 417, L33

\bibitem[{{Suyu} {et~al.}(2006){Suyu}, {Marshall}, {Hobson}, \&
  {Blandford}}]{suyu06}
{Suyu}, S.~H., {Marshall}, P.~J., {Hobson}, M.~P., \& {Blandford}, R.~D. 2006,
  \mnras, 371, 983

\bibitem[{{Swaters} {et~al.}(2003){Swaters}, {Madore}, {van den Bosch}, \&
  {Balcells}}]{swaters03}
{Swaters}, R.~A., {Madore}, B.~F., {van den Bosch}, F.~C., \& {Balcells}, M.
  2003, \apj, 583, 732

\bibitem[{{Treu} {et~al.}(2006){Treu}, {Koopmans}, {Bolton}, {Burles}, \&
  {Moustakas}}]{treu06}
{Treu}, T., {Koopmans}, L.~V., {Bolton}, A.~S., {Burles}, S., \& {Moustakas},
  L.~A. 2006, \apj, 640, 662

\bibitem[{{Treu} \& {Koopmans}(2002)}]{treu02}
{Treu}, T. \& {Koopmans}, L.~V.~E. 2002, \apj, 575, 87

\bibitem[{{Treu} \& {Koopmans}(2004)}]{treu04}
{Treu}, T. \& {Koopmans}, L.~V.~E. 2004, \apj, 611, 739

\bibitem[{{Trujillo} {et~al.}(2004){Trujillo}, {Burkert}, \&
  {Bell}}]{trujillo04}
{Trujillo}, I., {Burkert}, A., \& {Bell}, E.~F. 2004, \apjl, 600, L39

\bibitem[{{van Albada} {et~al.}(1985){van Albada}, {Bahcall}, {Begeman}, \&
  {Sancisi}}]{albada85}
{van Albada}, T.~S., {Bahcall}, J.~N., {Begeman}, K., \& {Sancisi}, R. 1985,
  \apj, 295, 305

\bibitem[{{Warren} \& {Dye}(2003)}]{warren03}
{Warren}, S.~J. \& {Dye}, S. 2003, \apj, 590, 673

\bibitem[{{Warren} {et~al.}(1996){Warren}, {Hewett}, {Lewis}, {Moller},
  {Iovino}, \& {Shaver}}]{warren96}
{Warren}, S.~J., {Hewett}, P.~C., {Lewis}, G.~F., {et~al.} 1996, \mnras, 278,
  139

\bibitem[{{Wayth} \& {Webster}(2006)}]{wayth06}
{Wayth}, R.~B. \& {Webster}, R.~L. 2006, \mnras, 372, 1187

\end{thebibliography}
\end{document}